\documentclass[floatfix,preprintnumbers,aps,prd,twocolumn,nofootinbib,amssymb,superscriptaddress,showkeys]{revtex4-1}

\usepackage{amsmath}
\usepackage{amsfonts}
\usepackage{amssymb}
\usepackage{hyperref}
\usepackage[dvipsnames]{xcolor}
\usepackage{graphicx}

\def\al{\alpha} 
\def\be{\beta} 
\def\ga{\gamma}

\def\ep{\epsilon}

\def\th{\theta}

\def\la{\lambda}

\def\ta{\tau}

\def\De{\Delta}

\def\pa{\partial}

\def\al{\alpha} 
\def\be{\beta} 
\def\ga{\gamma}

\def\ep{\epsilon}

\def\th{\theta}

\def\la{\lambda}

\def\ta{\tau}

\def\De{\Delta}

\def\pa{\partial}

\newcommand{\cs}{c_\text{s}} % Sound speed

\newcommand{\EnhFac}{f}

\newcommand{\HN}{H_\text{n}} % Hubble rate at nucelation
\newcommand{\Hn}{H_\text{n}}

\newcommand{\hstop}{h_\text{x}}
\newcommand{\estop}{e_\text{x}}

\newcommand{\nb}{n_\text{b}}
\newcommand{\Nb}{N_\text{b}}

\newcommand{\OmGW}{\Omega_\text{gw}}

\newcommand{\Rbc}{R_*} % Mean droplet radius at collision

\newcommand{\strPar}{\alpha_\text{n}}
\newcommand{\StrPar}{\alpha_\text{n}}

\newcommand{\tShock}{\ta_\text{sh}}

\newcommand{\Tfluid}{T^\text{f}}

\newcommand{\tLife}{\tau_\text{v}}
\newcommand{\TN}{T_\text{n}} % Nucleation temperature
\newcommand{\Tn}{T_\text{n}} % Nucleation temperature
\newcommand{\tn}{t_\text{n}} % Nucleation time
\newcommand{\Tc}{T_\text{c}} % Critical temperature
\newcommand{\tc}{t_{\text{c}}} % Critical time
\newcommand{\Tu}{\bar T}

\newcommand{\tstop}{t_\text{x}}
\newcommand{\taustop}{\tau_\text{x}}

\newcommand{\vw}{v_\text{w}} % Wall speed
\newcommand{\vweff}{v_\text{eff}} % Wall speed
\newcommand{\vShock}{v_\text{sh}}
\newcommand{\vsh}{v_\text{sh}} % Shock speed
\newcommand{\Vu}{V}
\newcommand{\Vb}{V_\text{s}}
\newcommand{\Veff}{V_\text{eff}}
\newcommand{\Vbeff}{V_\text{s,eff}}
\newcommand{\Vtot}{V_\text{tot}}
\newcommand{\vip}{v_\text{ip}}
\newcommand{\vmax}{v_\text{max}}

\definecolor{newgreen}{RGB}{10,100,20}

\newcommand{\ben}{\begin{equation}}
\newcommand{\een}{\end{equation}}
\newcommand{\bea}{\begin{eqnarray}}
\newcommand{\eea}{\end{eqnarray}}
\newcommand{\ba}{\begin{array}}
\newcommand{\ea}{\end{array}}
\newcommand{\bit}{\begin{itemize}}
\newcommand{\eit}{\end{itemize}}

\newcommand{\nn}{\nonumber}

\newcommand{\bx}{\textbf{x}}

\begin{document}

\newcommand{\Sussex}{\affiliation{
Department of Physics and Astronomy,
University of Sussex, Falmer, Brighton BN1 9QH,
U.K.}}

\newcommand{\HIPetc}{\affiliation{
Department of Physics and Helsinki Institute of Physics,
PL 64, % (Gustaf H\"{a}llstr\"{o}min katu 2),
FI-00014 University of Helsinki,
Finland
}}

\title{Thermal suppression of bubble nucleation at first-order phase transitions in the early Universe}

\author{Mudhahir Al Ajmi} \email{mudhahir@squ.edu.om}
\affiliation{{Department of Physics, College of Science, Sultan Qaboos University,
		\\P.O. Box 36, P.C. 123,  Muscat, Sultanate of Oman}}
\author{Mark Hindmarsh}
\email{mark.hindmarsh@helsinki.fi}
\HIPetc
\Sussex

\date{\today}

\begin{abstract}
One of the key observables in a gravitational wave power spectrum from a first order phase transition in the early Universe is the mean bubble spacing, which depends on the rate of nucleation of bubbles of the stable phase, as well as the bubble wall speed.  When the bubbles expand as deflagrations, it is expected that the heating of the fluid in front of the phase boundary suppresses the nucleation rate.  We quantify the effect,  showing that it increases the mean bubble separation,  and acts to enhance the gravitational wave signal by a factor of up to order 10. 
The effect is largest for small wall speeds and strong transitions. 
\end{abstract}
\preprint{HIP-2022-3/TH}

\maketitle

\section{Introduction}

An early Universe cosmological first-order phase transition \cite{Steinhardt:1981ct} 
can lead to interesting physical consequences such as matter-antimatter asymmetry \cite{Kuzmin:1985mm}, 
primordial magnetic fields \cite{Hogan:1983zz,Baym:1995fk}
and the production of a stochastic background of 
gravitational waves \cite{Witten:1984rs,1986MNRAS.218..629H} 
The power spectrum of the gravitational waves contains information about the thermodynamic and transport properties of the 
system at the time of the phase transition. If the transition happened at around the electroweak scale of 100 GeV, when the Universe was about $10^{-11}$ seconds old,  
the gravitational waves could be observable at planned space-based detectors like Laser Interferometer Space Antenna (LISA) \cite{LISA:2017pwj}, 
\cite{Caprini:2019egz}, 
and the principal thermodynamic and transport properties could be measured over a wide region of the parameter space \cite{Gowling:2021gcy}. 

The transformation of the metastable phase into the stable one is described by cosmological homogeneous nucleation theory 
\cite{Steinhardt:1981ct,Linde:1981zj} (see \cite{Hindmarsh:2020hop} for a review). 
Once the temperature has fallen below the critical temperature of the transition $\Tc$,  
quantum or thermal fluctuations produce small spherical bubbles of the stable phase, 
which expand and merge, and eventually the whole Universe is converted to its stable phase. 
The peak of the bubble nucleation rate defines the nucleation temperature $\Tn$, and the time taken 
to complete the transition can be expressed as a transition rate $\beta$. 
Part of the potential energy of the supercooled metastable phase is converted into bulk fluid motion, which 
sources gravitational waves long after the transition is completed \cite{Hindmarsh:2013xza}. 
The kinetic energy of the bulk fluid motion is controlled by a fourth parameter $\alpha$, which is essentially the ratio of 
the potential energy difference to the thermal energy. It quantifies the strength of the transition. 
The kinetic energy of the bulk motion is also controlled by the equation of state of the fluid, most importantly 
through the speeds of sound in the two phases \cite{Giese:2020znk}.
All quantities will be defined more precisely below.

At the end of the transition, the density of bubble nucleation sites $n_*$ defines a mean bubble spacing $\Rbc = 1/n_*^{1/3}$. 
If bubble nucleation continues undisturbed in the metastable phase, and the bubbles expand at a constant speed $\vw$, 
the mean bubble spacing is given by 
\ben
\Rbc = (8\pi)^{1/3} \vw/\beta.
\een
However, if the bubbles expand as deflagrations \cite{LanLifFlu, Steinhardt:1981ct, KurkiSuonio:1995vy, Espinosa:2010hh, Hindmarsh:2019phv}, the fluid ahead of the advancing bubble wall is heated, out to the radius of a shock.
In this region the nucleation rate is reduced. Thus we expect fewer bubbles to be nucleated, and the mean bubble spacing should increase. 
The effect acts to increase the peak wavelength of the gravitational wave power spectrum \cite{Hindmarsh:2016lnk,Hindmarsh:2017gnf}.
and also the peak power, which scales as $R_*$ or $R_*^{2}$, depending on the strength of the transition 
\cite{Caprini:2019egz}. 

The suppression of nucleation by the heating effect has been noted before \cite{Moore:2000jw}, where it was estimated that nucleation was suppressed everywhere between the wall and the shock.  In this paper we quantify the effect more precisely, for transitions with strength parameters up to $\alpha \simeq 1$. 
We find that the increase in the mean bubble spacing can increase the gravitational wave power 
by a factor of up to O(10), partially 
compensating the suppression due to the interactions between bubbles \cite{Cutting:2019zws}.     

\section{Bubble nucleation: standard calculation}

First, we discuss the formation of bubbles in the plasma according to the standard treatment.  
We will suppose that the transition rate is much faster than the Hubble rate $H$, so that 
we can neglect cosmic expansion.
%\subsection{Bubble nucleation}
The bubble nucleation rate per unit volume has the form \cite{Guth:1981uk,Linde:1981zj,Enqvist:1991xw,Hindmarsh:2019phv}
\ben
p(t) = p_0 e^{-S(T(t))}.
\een
where $S$ is the action for the appearance of a bubble, which in a thermal transition is equal to the energy of a  
critical bubble divided by the temperature. It is a function of time through its dependence on the temperature $T(t)$. 
In the small supercooling or ``thin wall'' approximation \cite{Linde:1981zj,Enqvist:1991xw}, 
\ben
\label{e:BubAct}
S \simeq \frac{s_0}{|(1 - \hat T)]^2},
\een
where  $\hat T = T/\Tc$, $\Tc$ is the critical temperature, and $s_0$ is a constant computable from the 
effective potential of the theory.
The nucleation rate is then zero precisely at the critical temperature, and increases very rapidly below it. 

Once a bubble has nucleated, it grows at a constant speed $\vw$, determined by the friction between the wall and the plasma.
The increasing population of growing bubbles reduces the fraction of the Universe in the metastable phase. 

Let $\Vu$ be the volume in the metastable phase, and $\Vb$ the volume in the stable phase, out of a total volume $\Vtot$, such that 
$$
\Vtot = \Vu + \Vb.
$$
In the notation of Ref.~\cite{Enqvist:1991xw}, 
\ben
h = \Vu/\Vtot
\een
denotes the fraction of the Universe in the metastable (high-temperature) phase. 
First, we consider the reduction in the volume of the unbroken phase between times $t$ and $t+dt$ due to the growth of bubbles nucleated between earlier times $t'$ and $t'+dt'$:
\ben
d^2\Vu(t,t') = - d\Nb(t') 4\pi {R}^2 dR \frac{V(t)}{V(t')},
\een
where $d\Nb$ is the number of bubbles nucleated in that time interval, and $R$ is the radius of those bubbles at time $t$.  The factor $V(t)/V(t')$ takes into account the fact that only parts of the bubbles growing into the unbroken phase will change the volume of that phase.

The number of bubbles nucleated between $t'$ and $t'+dt'$ is 
\ben
d\Nb = p(t') V(t') dt',
\een
where $p(t')$ is the bubble nucleation rate per unit volume, and 
the factor $V(t')$ accounts for the fact that bubbles nucleate only 
in the metastable phase.  
Finally, we have 
\ben
R = \vw(t - t'), \;\; dR = \vw dt, 
\een
as the bubbles are assumed to grow with constant speed after nucleation. 

The nucleation probability is non-zero only below the critical temperature $\Tc$, which is reached at time $\tc$, so the change in the volume of the stable phase between $t$ and $t+dt$ is, in total,
\ben
d\Vu(t) = -\vw\Vu(t) dt \int_{\tc}^{t} dt' p(t') 4\pi \vw^2(t - t')^2. 
\een
Dividing by the total volume $\Vtot$, we obtain a differential equation for $h$, the fraction remaining in the metastable phase:
\ben
\label{e:hDE}
\frac{dh}{dt} = -\vw h(t) \int_{\tc}^{t} dt' p(t') 4\pi \vw^2(t - t')^2. 
\een
It is straightforward to check that the solution to this equation, with the boundary condition $h(t) = 1$ for $t < \tc$, is 
\ben
h(t) = \exp\left( -  \frac{4\pi}{3}\int_{\tc}^{t} dt' p(t')  \vw^3(t - t')^3 \right).
\een
A saddle-point approximation to the integral is possible. 
We define the transition rate parameter 
\ben
\beta = \left. \frac{d }{dt} \ln p(t) \right|_{t_f} ,
\een
where $t_f$ will be specified later, 
and write
\begin{equation}
\label{e:ProApp}
p(t) \simeq p_f e^{\beta(t-t_f)}
\end{equation}
where $p_f = p_0 \exp(-S(t_f))$.
The integral can be performed in the approximation $\beta(\tc - t_f) \to - \infty$, giving
\ben
h(t) = \exp\left( -  {8\pi} \vw^3 \be^{-4}p_f e^{\be t} \right).
\een
Choosing $t_f$ to be the time at which $h(t_f) = 1/e$, 
we have
\ben
h(t) = \exp\left(- e^{\be(t - t_f)} \right) .
\een
with
\ben
{8\pi} \vw^3 \be^{-4} p_f = 1 .
\een
The time $t_f$ is then found as the solution to 
\ben
{8\pi} \vw^3 \be^{-4} p_0 e^{- S(t_f)} = 1 .
\een
Its value does not play an essential role in the following.

To calculate the number of bubbles, we integrate the equation
\ben
\frac{d\Nb}{dt'} =  p(t')  V(t')  ,
\een
or, in terms of the bubble density $\nb = \Nb/\Vtot$,
\ben
\frac{d\nb}{dt'} =  p(t')h(t') .
\een
The final bubble density is then
\ben
\nb =  \be^{3}/8\pi\vw^3 .
\een
We define a mean bubble centre spacing $\Rbc$ as 
\begin{equation}
\Rbc = \nb^{-1/3} = (8\pi)^{1/3} (\vw/\be) .
\end{equation}
Note that we have assumed that the bubble nucleation rate 
is the same everywhere in the metastable phase.  However, 
the expanding bubble releases energy and heats up the fluid, as well as setting it in motion. 
If this heating effect extends in front of the bubble wall, as it does for 
deflagrations, we can see from (\ref{e:BubAct}) that the nucleation rate will be reduced. \footnote{Temperature fluctuations from other sources can also change the nucleation rate \cite{Jinno:2021ury}. }
In the next section we review the calculation the temperature profile around an expanding bubble.

\section{Hydrodynamics of bubble growth by deflagrations}

In order to calculate the suppression effect noted in the previous section, we need to calculate the temperature profile around an expanding bubble for deflagrations.  

In the case of perfect fluid the plasma is locally in equilibrium. Therefore, the energy momentum tensor of the fluid can be written as:
\ben
\label{eq:Tmunu}
\Tfluid_{\mu\nu} = w \, u_\mu u_\nu  - g_{\mu\nu} \, p, 
\een
where $w$ is the enthalpy density and $p$ is the pressure.
The four-velocity field of the plasma $u^\mu$ is related to its
three-velocity $\mathbf{v}$ by
\ben
u^\mu = \frac{(1, \mathbf{v} )}{\sqrt{1-\mathbf{v}^2}}
= (\gamma, \gamma\mathbf{v} ) \ .
\een
The energy density, $e$, the enthalpy density $w$, the entropy density $s$ are related to the pressure as follows:
\ben
\label{eq:enthsige}
e\equiv T\frac{\partial p}{\partial T} -p\ ,\quad
w \equiv T\frac{\partial p}{\partial T}\ ,\quad
s \equiv \frac{\partial p}{\partial T}\ .
\een
Evidently,
$
w = e + p\ .
$
The motion of the fluid is governed by conservation of energy-momentum:
\ben
\label{eq:T_cons}
\partial^\mu \Tfluid_{\mu\nu} = 0 \ .
\een
The fluid around an expanding spherical bubble at time since nucleation $\Delta t$ and radial distance $r$ can be described by the purely radial fluid 3-velocity $v(r, \Delta t)$ and enthalpy $w(r, \Delta t)$. 
The partial differential equations can be solved numerically \cite{KurkiSuonio:1995vy}, and it is observed that 
the fluid quickly settles down to a self-similar solution,  depending only on a coordinate 
%Defining 
$\xi = r/\Delta t$. In this self-similar form, 
the fluid equations become 
\bea
	\frac{d v}{d \xi} &=& \frac{2v(1-v^2)}{\xi(1-\xi v)} \left(\frac{\mu^2}{c_{\rm s}^2}-1\right)^{-1}\,, \label{e:SelSimVel}\\
	\frac{d w}{d \xi} &=& w \left(1+\frac{1}{c_{\rm s}^2}\right)\gamma^2 \mu \frac{d v}{d \xi}\,. \label{e:SelSimEnt}
\eea
Here, $c_{\rm s}^2=d p/d e$ is the speed of sound and
\ben
	\mu=\frac{\xi-v}{1-\xi v}.
\een
is the fluid velocity at $\xi$ in a frame that is moving outward at speed $\xi$. 

We now discuss the boundary conditions at the bubble wall. 
The plasma is in the stable phase behind the wall, and in the metastable phase in front. We denote quantities evaluated just behind the wall with a subscript $-$, and in front with a subscript $+$, and quantities in the frame moving with the wall with a tilde.

In the frame moving with the wall, conservation of energy density and momentum density imply
\begin{eqnarray}
    w_+ \tilde\gamma_+^2 + p_+ &=& w_- \tilde\gamma_-^2 + p_- ,\\
    w_+ \tilde\gamma_+^2 \tilde{v}_+ &=& w_- \tilde\gamma_-^2 \tilde{v}_- ,
\end{eqnarray}
These equations may be rearranged to give 
\bea
\label{eq:vvs2}
\tilde{v}_+ &=& \frac{1}{1+\alpha_+}\left[ \left(\frac{\tilde{v}_-}{2}+\frac{1}{6
\tilde{v}_-}\right)\right. \nn\\
&\pm& \left.\sqrt{\left(\frac{\tilde{v}_-}{2}+\frac{1}{6 \tilde{v}_-}\right)^2 +
\alpha_+^2 +\frac{2}{3}\alpha_+ - \frac{1}{3}} \right], 
\eea
where
\ben
\label{alphar}
\al_+=\frac{4}{3}\frac{\theta_+-\theta_-}{w_+}\ ,
\een
and $\theta = (e - 3p)/4$. 
In a deflagration, the fluid exits from the wall with the smaller of the wall speed and the sound speed in the stable phase, 
\begin{equation}
    \tilde{v}_- = \min(\vw, \cs(T_-))
\end{equation}

These equations can be straightforwardly solved numerically, once the speed of sound is known as a function of enthalpy density.  In practice one wishes to apply the boundary condition at large radii, where $v = 0$ and $T = \Tn$, the bubble nucleation temperature.  
This means that one must ``shoot'' for this value from an initial guess for $\alpha_+$. The amount of energy available to be released by the transition is conveniently parametrised by 
\ben
\strPar = \frac{4}{3}\frac{\theta_\text{m}(\Tn) - \theta_\text{s}(\Tn)}{w_\text{m}(\tn)},
\een
where the subscripts m and s denote quantities evaluated in the metastable and stable phases. 

In this paper we use the simplest possible equation of state, the bag model, in which 
\ben{\label{e:BagEOS}}
p_\text{m} = \ep + a_\text{m} T ^4, \quad
p_\text{s} = a_\text{s} T ^4 .
\een
In this model $\cs^2 = 1/3$ and $w \propto T^4$.  

\begin{figure}[htb!]
\centering
\includegraphics[width=0.4\textwidth]{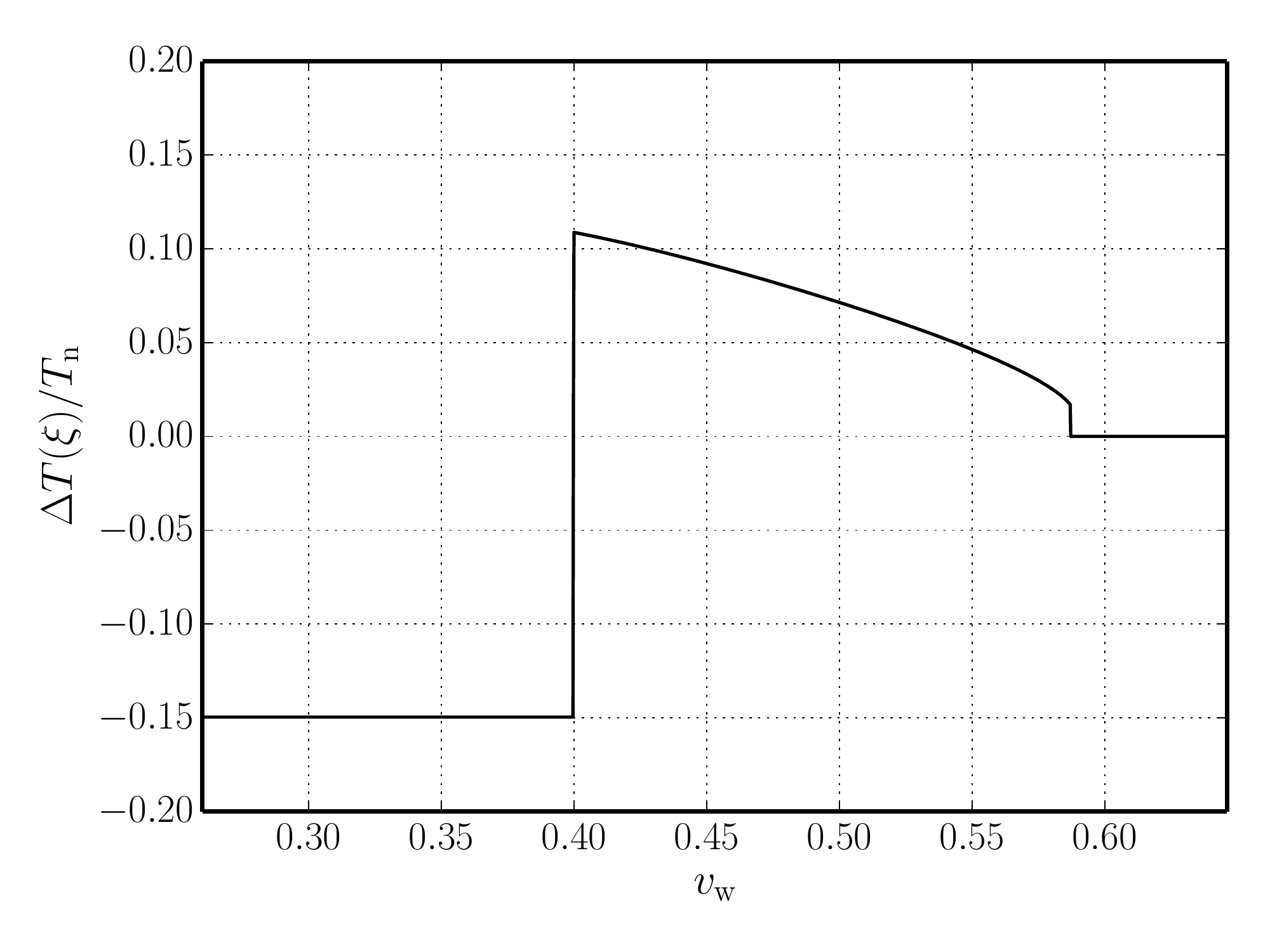}\\
\includegraphics[width=0.4\textwidth]{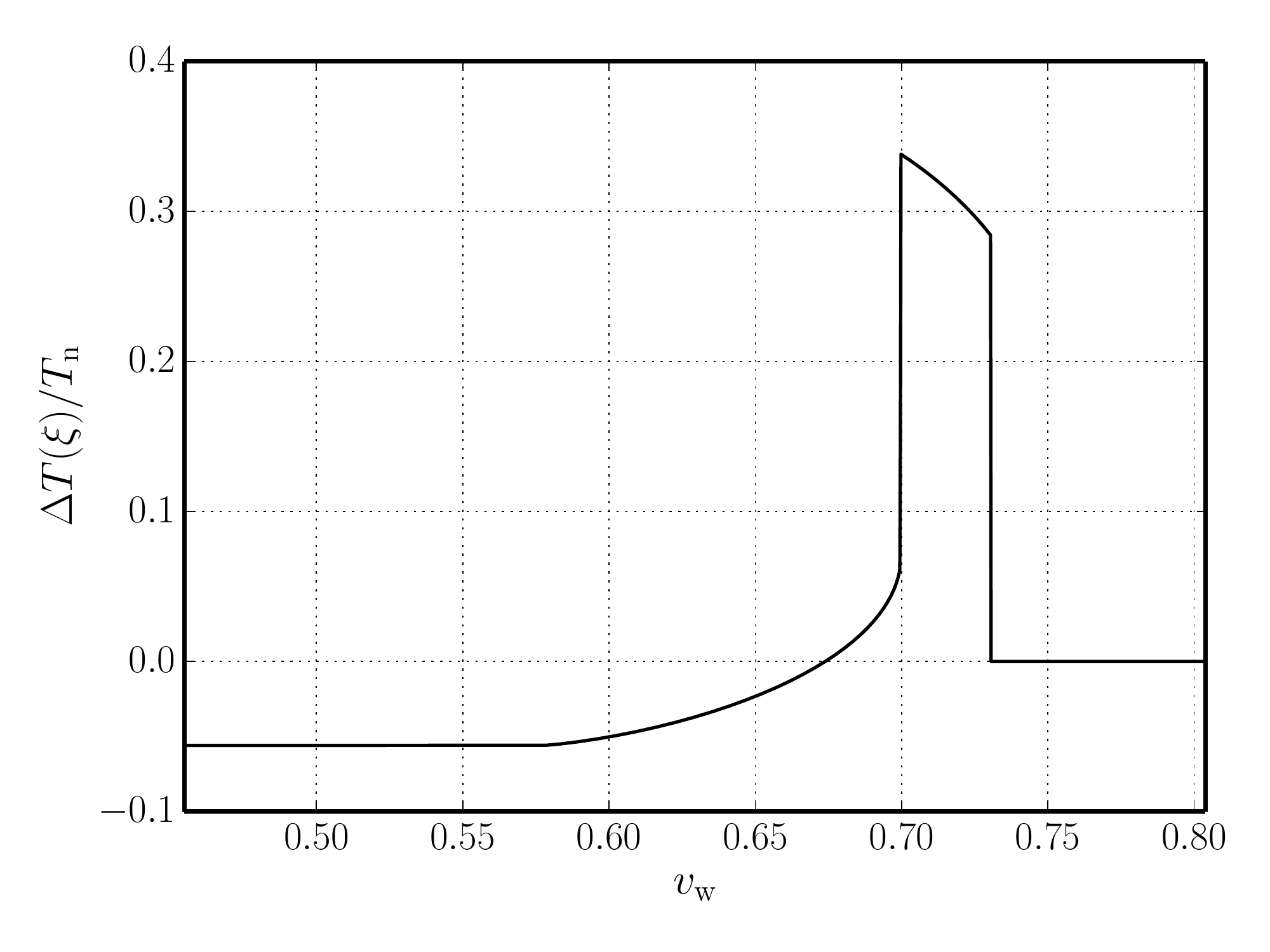}%\par\medskip
\caption{Variation of temperature with respect to radial similarity variable $\xi = r/t$ in the frame of the bubble center for different wall velocities, $\vw=0.4$, leading to a deflagration (top) and $\vw = 0.7$ leading to a hybrid (bottom), and phase transition strength parameter $\StrPar = 0.25$. The plots show the typical profiles for a deflagrations which are subsonic (top) and supersonic  (bottom).
}  
\label{temeratureprofile}
\end{figure}

Figure \ref{temeratureprofile} shows plots of temperature profile variation with respect to $\xi$ for different wall speeds  ($\xi=0.4,0.7$) and for $\strPar=0.2$. 
The plots were produced by integrating the equations (\ref{e:SelSimVel}, \ref{e:SelSimEnt}) in parametric form  (see \cite{Hindmarsh:2019phv}), with 5000 points distributed over the parameter range, and an absolute tolerance of $10^{-6}$ in the shooting parameter $\alpha_+$. 

The upper plot shows the typical form of a subsonic deflagration and the lower one shows a supersonic deflagration (or hybrid) \cite{KurkiSuonio:1995pp}. 
In each case there is a region of heated fluid, which is moving radially outward, bounded on the outside by a shock, which moves at a speed $\vsh$ calculable from the solution to the fluid equations $v(\xi)$ and energy-momentum conservation (see e.g.~\cite{Hindmarsh:2019phv}.  The resulting equation gives the position of the shock as the solution to the equation 
\begin{equation}
    v(\xi) = \frac{3\xi^2 - 1}{2\xi} .
\end{equation}
Between the bubble wall and the shock the bubble nucleation rate will be suppressed.

\begin{figure}[h!]
\centering
\includegraphics[width=0.4\textwidth]{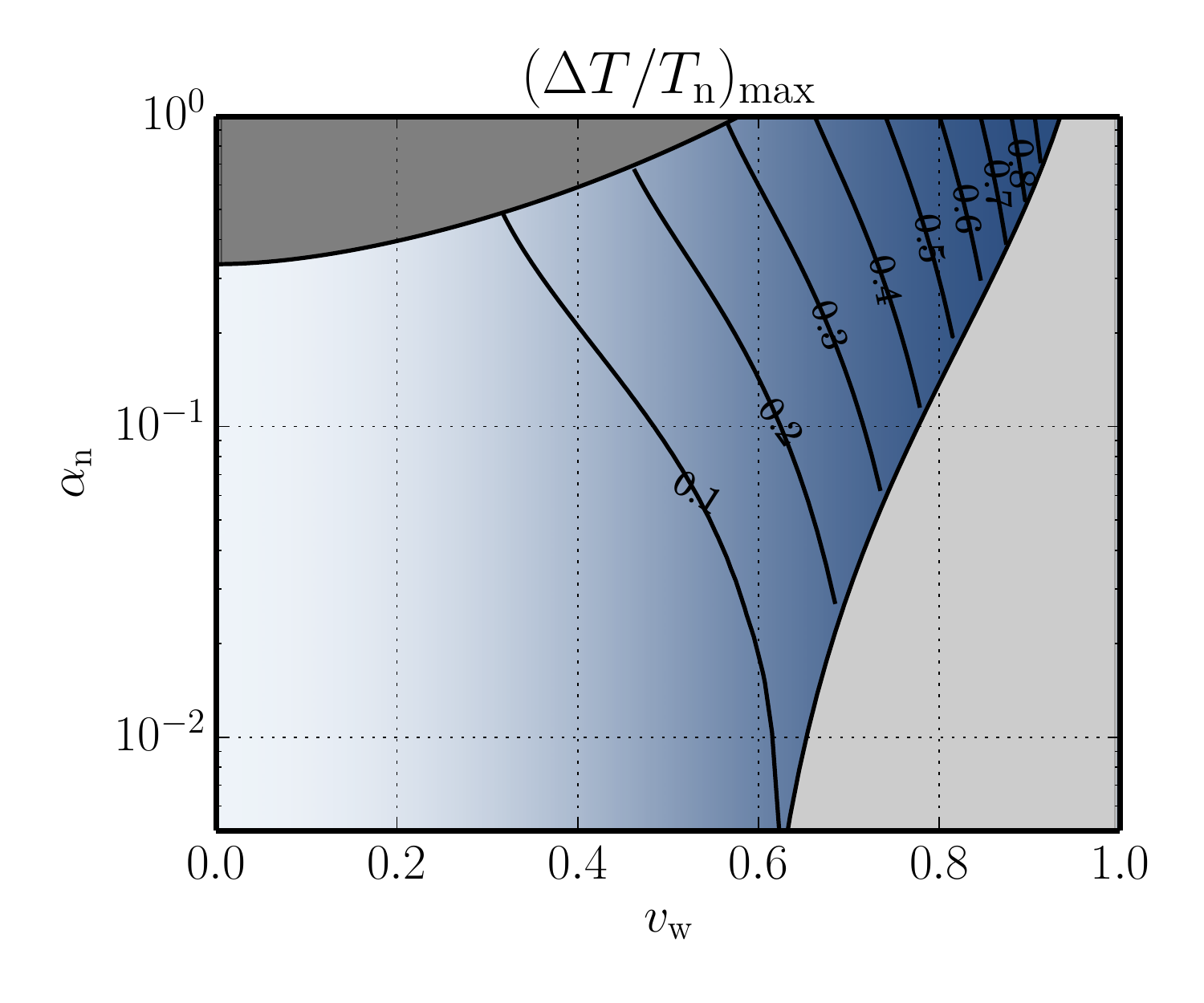}\\  %\hfil
\caption{Contours of the maximum fractional temperature difference around a deflagration, as a function of 
wall speed $v_w$ and strength parameter at the nucleation temperature $\StrPar$. 
The blue shading also shows the maximum fractional temperature difference, for use in other figures.
For larger values of $\vw$, deflagrations are replaced by detonations and there is no effect.  For larger values of $\StrPar$, there is no 
hydrodynamic solution at fixed $\vw$.  Both regions are greyed out. 
}
\label{contourtempdiff}
\end{figure}

Figure \ref{contourtempdiff} shows the maximum of the ratio of the  temperature difference $\Delta T = T - \Tn$ to the nucleation temperature $\Tn$ as contour lines in the plane of wall speed $\vw$ and transition strength $\StrPar$, 
for a range of $\vw=0.01-0.99$ and $\strPar=0.005-1.0$. 
We also show $\Delta T/\Tn$ with blue shading, and use it in later figures to give an qualitative indication of the accuracy of the calculations, 
from light (more accurate) to dark blue (less accurate).
% Towards the detonation region the ratio increases. At small $v_w$ the values of the ratio are small.
For large $\StrPar$ and low $\vw$ there are no solutions to the fluid equations; for large $\vw$ at fixed $\StrPar$ the solution changes to a detonation, where the wall moves ahead of the shock, and the nucleation suppression effect disappears.

\subsection*{Small wall speed limit}

In the limit of small wall speed, the fluid velocity should be small $(v(\xi) \ll \xi)$, and a simple approximation to the solution to the fluid equations is known. 
The approximation is
\ben
\label{e:Vip}
\vip(\xi) = \vmax \frac{\vw^2}{\xi^2}\frac{\cs^2- \xi^2}{\cs^2-\vw^2},
\een
where 
$\vmax(\vw,\al)$ is 
the maximum fluid speed in the bubble centre frame, reached just outside the wall.
For small $\al$ and $\vw$ not too close to the sound speed \cite{Espinosa:2010hh,Hindmarsh:2020hop},  
\begin{equation}
\vmax \simeq 3\strPar\vw \frac{1}{1 - 3\vw^2} .
\end{equation}
This gives an analytic form for the enthalpy,
\ben
w(\xi) \simeq w_\text{n} \exp\left[ 2(1+\cs^2) \frac{\vw^2}{\cs^2 - \vw^2} \vmax \left( \frac{1}{\xi} - \frac{1}{\cs} \right)  \right]
\een
where $w_\text{n}$ is the enthalpy density at $\xi \ge \cs$, the enthalpy density at the nucleation temperature.

Because $w \propto T^4$, we have for small enthalpy differences $\Delta w/w = 4 \Delta T/T$. 
Hence 
\bea
\frac{\Delta T}{\TN} &\simeq& \frac{1}{2}(1+\cs^2) \frac{\vw^2}{\cs^2 - \vw^2} \vmax \left( \frac{1}{\xi} - \frac{1}{\cs} \right) \nonumber \\
&\simeq& \frac{3\strPar}{2}(1+\cs^2) \frac{\vw^3 }{\cs^2(1 - 3\vw^2)^2}   \left( \frac{1}{\xi} - \frac{1}{\cs} \right) \nonumber \\
&\simeq& {6\strPar} \frac{\vw^3 }{(1 - 3\vw^2)^2}   \left( \frac{1}{\xi} - \sqrt{3} \right) 
\label{e:delTEq}
\eea
This relation is useful in the low velocity region when $v \lesssim 0.05$.

\section{Position-dependent bubble nucleation}

In this section, we take into account the dependence of the bubble nucleation rate on the temperature. The temperature is raised closer to the critical temperature in front of a deflagration, and so we expect the nucleation rate to be suppressed around the expanding bubble.

The bubble nucleation rate per unit volume is now space-dependent as well as time dependent, due to the dependence of the bubble appearance action on the temperature:
\ben
p(t',\bx) = p_0 e^{-S(t', \bx)}.
\een
Let us write
\ben
T(t',\bx) = \Tu(t') + \De T(t', \bx),
\een
where $\Tu$ is the undisturbed temperature outside the shock surrounding each bubble. Hence
\ben
p(t',\bx) = p(t') e^{-\De S(t',\bx)}
\een
The rate of bubble nucleation over the remaining volume remaining in the metastable phase $V' \equiv V(t')$ is
\ben
\frac{d\Nb}{dt'} = \int_{V'} d^3 x \, p(t',\bx) ,
\een
Hence
\begin{align}
\frac{d\Nb}{dt'} %&= p(t')\int_{V'} d^3 x e^{-\De S} \\ 
&= p(t')\left( V' - \int_{V'} d^3 x ( 1 - e^{-\De S}) \right) ,
\label{e:BubNumDen}
\end{align}
where we have rearranged the equation to bring out a term which acts to reduce the effective volume of the metastable phase. 
Let us give the correction term the symbol
\ben
\label{e:DelVbeff}
\Delta\Vbeff =   \int_{V'} d^3 x ( 1 - e^{-\De S}).
\een
We can regard this quantity as an increase in the effective volume of the stable phase, where no bubbles can nucleate. 

Let us first consider a single bubble nucleated at time $t_1$, with radius $R = \vw(t' - t_1)$.  We can then write 
\ben
\Delta\Vbeff = \Vb \EnhFac,
\een
where 
\ben
\EnhFac = \frac{3}{\vw^3}\int_{\vw}^{\vShock} \xi^2 d\xi ( 1 - e^{-\De S})
\een
is a constant factor giving the relative increase in the effective volume of the bubble, 
$\vShock$ is the speed of the shock which surrounds the bubble, and the volume in the stable phase inside the bubble is $\Vb = 4\pi\vw^3(t' - t_1)^3/3$. 
The effective volume in the stable phase is then
\begin{equation}
    \Vbeff = (1 + f) \Vb .
\end{equation}
The change in the bubble appearance action $\Delta S$ can be expressed in terms of the transition rate parameter $\beta = -\pa S/\pa t$,
\ben
\De S \simeq \frac{\pa S}{\pa T} \De T = \frac{\pa S}{\pa t} \frac{\pa t}{\pa T} \De T = \frac{\beta}{H} \frac{\De T}{T}.  
\een
We neglect higher orders in ${\De T}/{T}$, which means that in the region of $(\vw,\strPar)$ plane where $\De T/T \sim 1$, 
our calculations can be expected to receive large corrections. These corrections will depend on the second and higher derivatives 
of the action with respect to the logarithm of the temperature, parameters which are not part of the standard description of phase transitions. We leave more precise studies for future work, indicating where corrections are likely to be more important by the density of the blue shading in the $(\vw,\strPar)$ plane.

\begin{figure*}[htbp]
\centering
\includegraphics[width=0.33\textwidth]{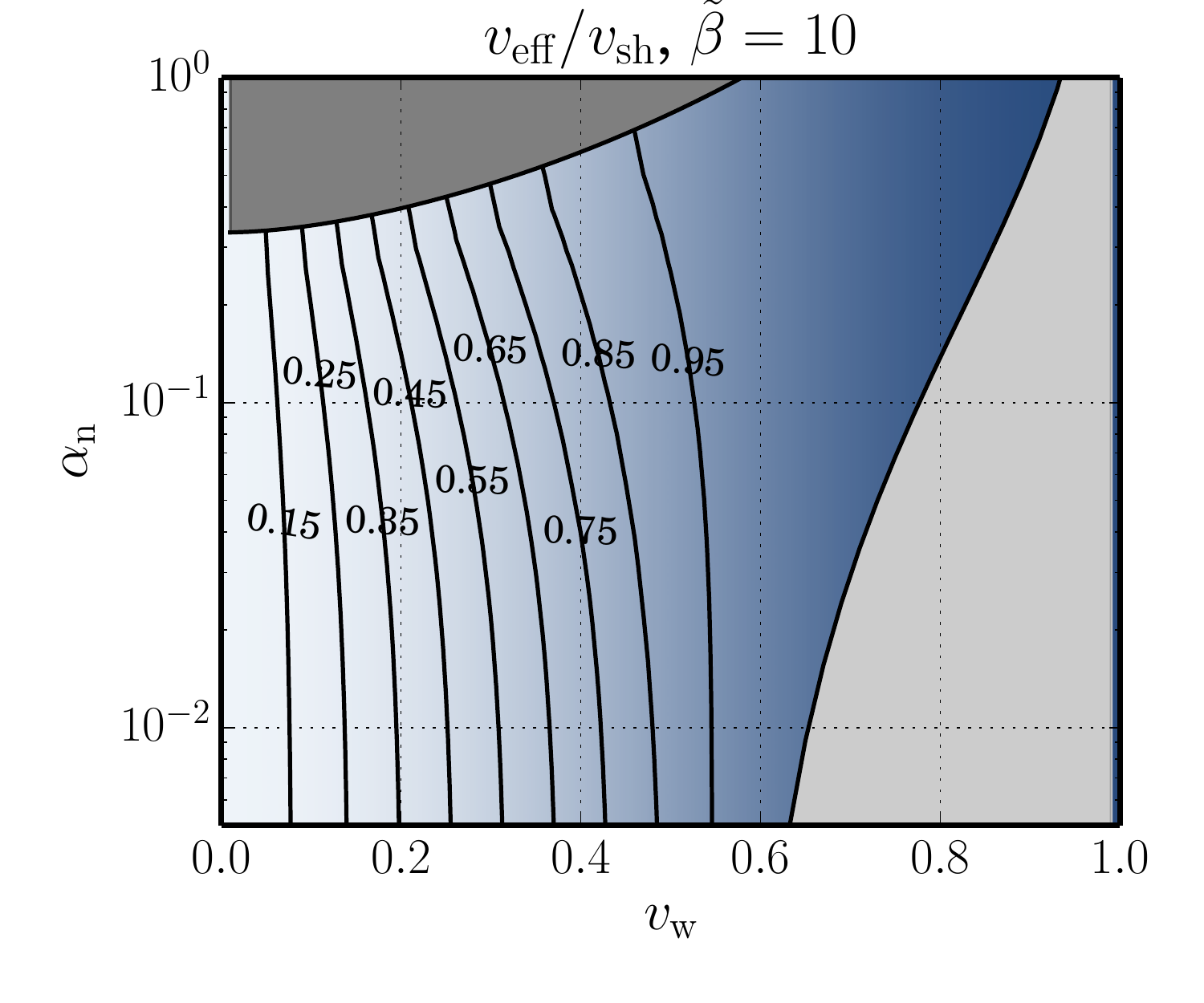}%\hfil
\includegraphics[width=0.33\textwidth]{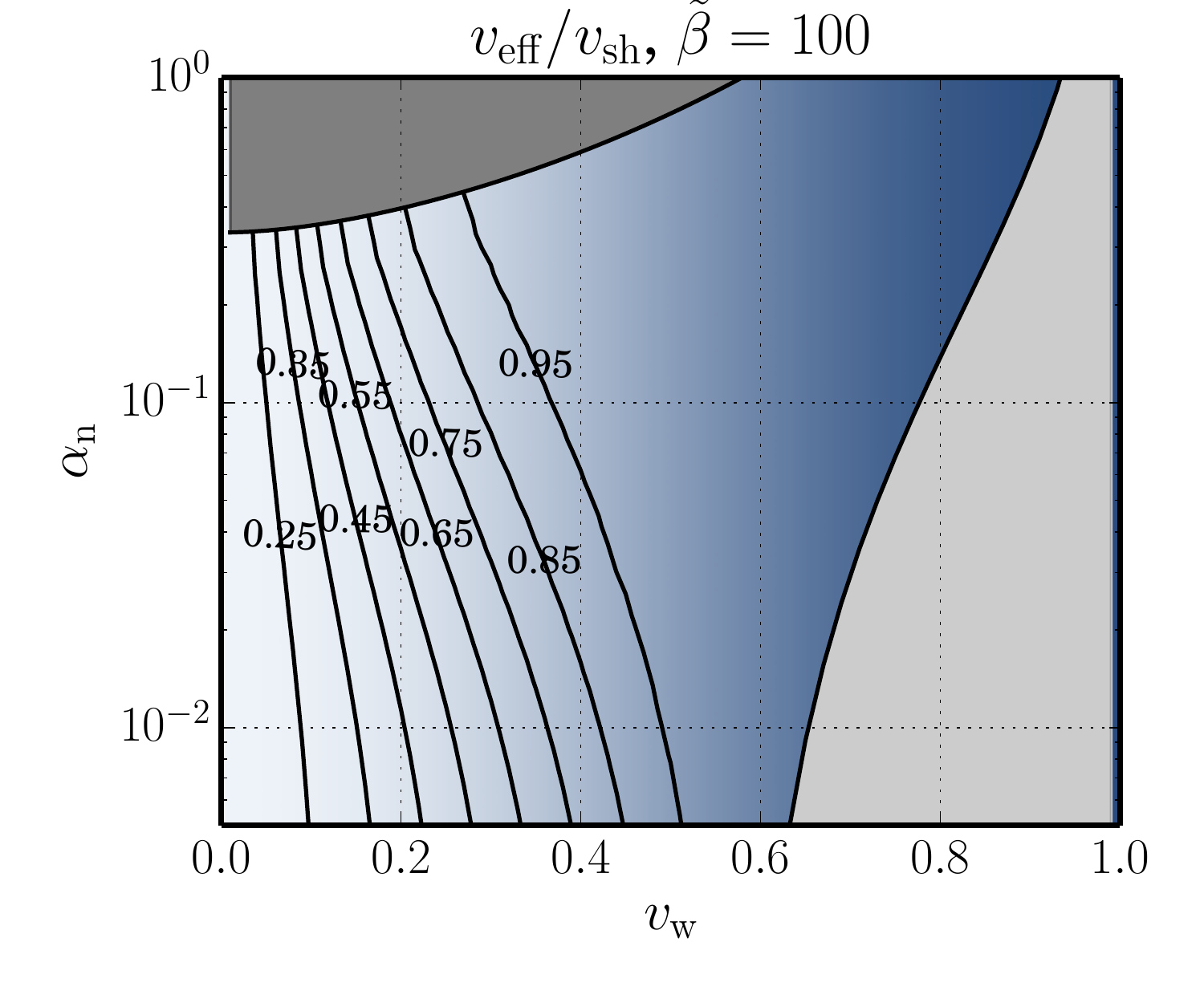}%\par\medskip
\includegraphics[width=0.33\textwidth]{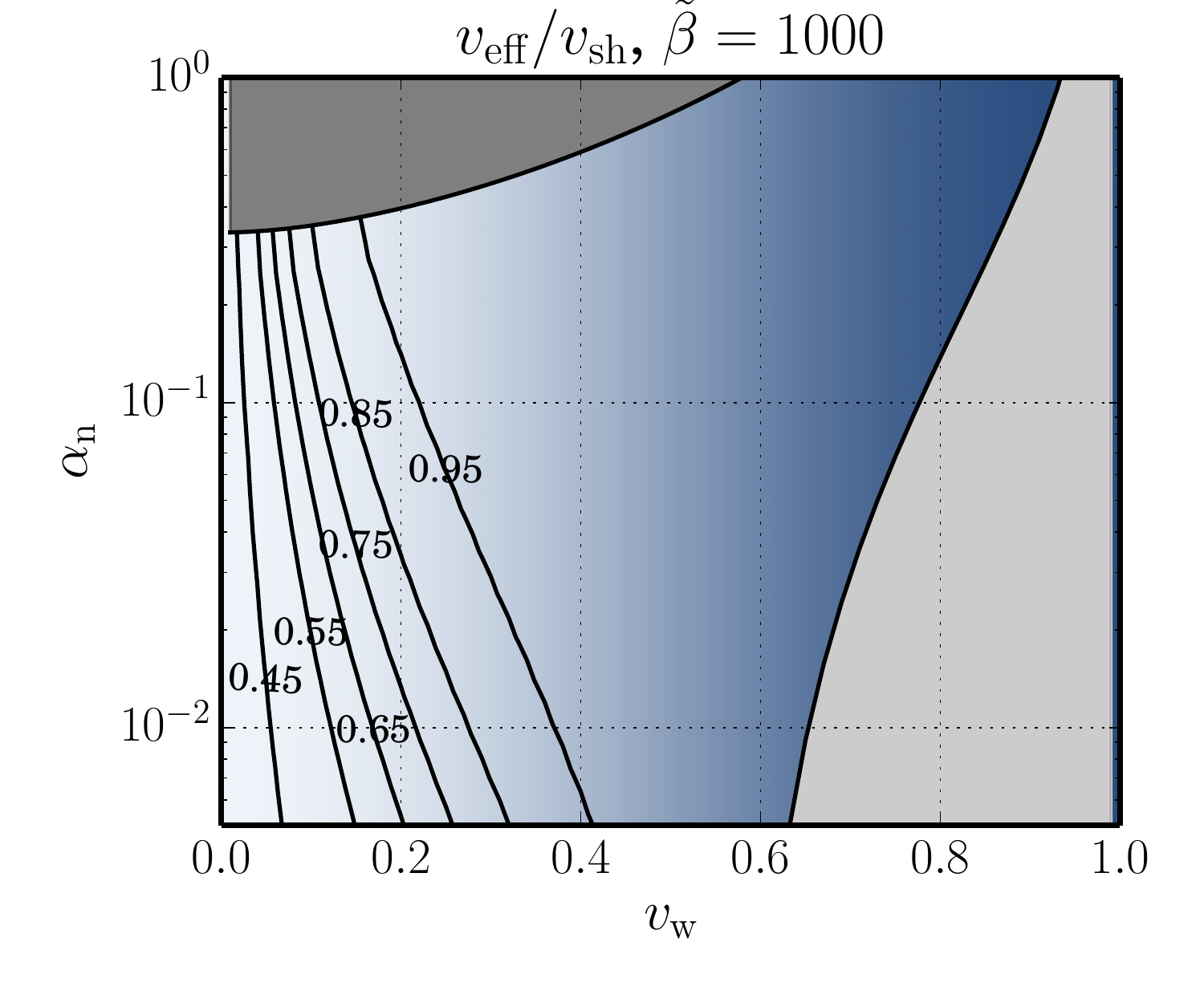}\\%\hfil
\includegraphics[width=0.33\textwidth]{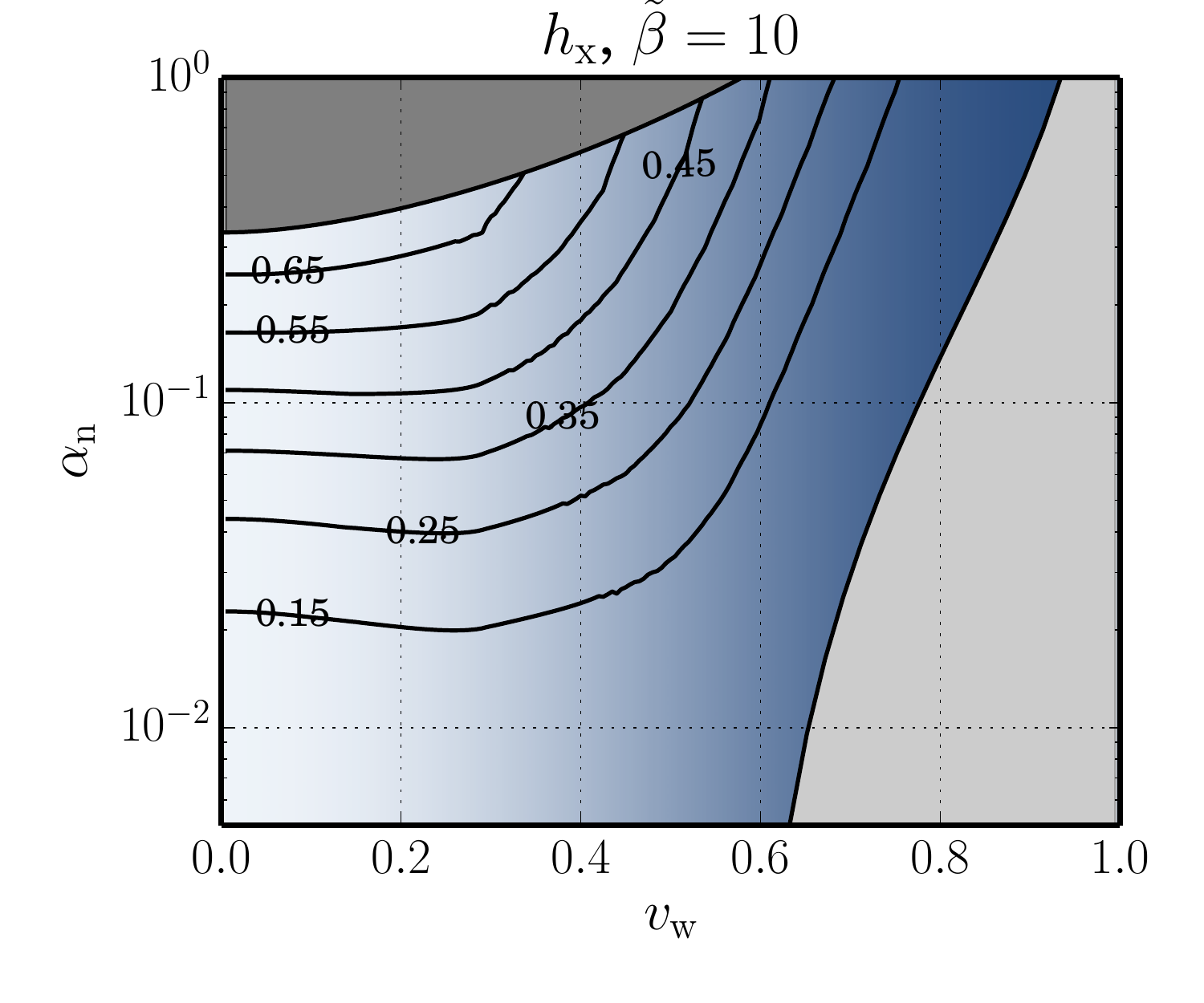}%\hfil
\includegraphics[width=0.33\textwidth]{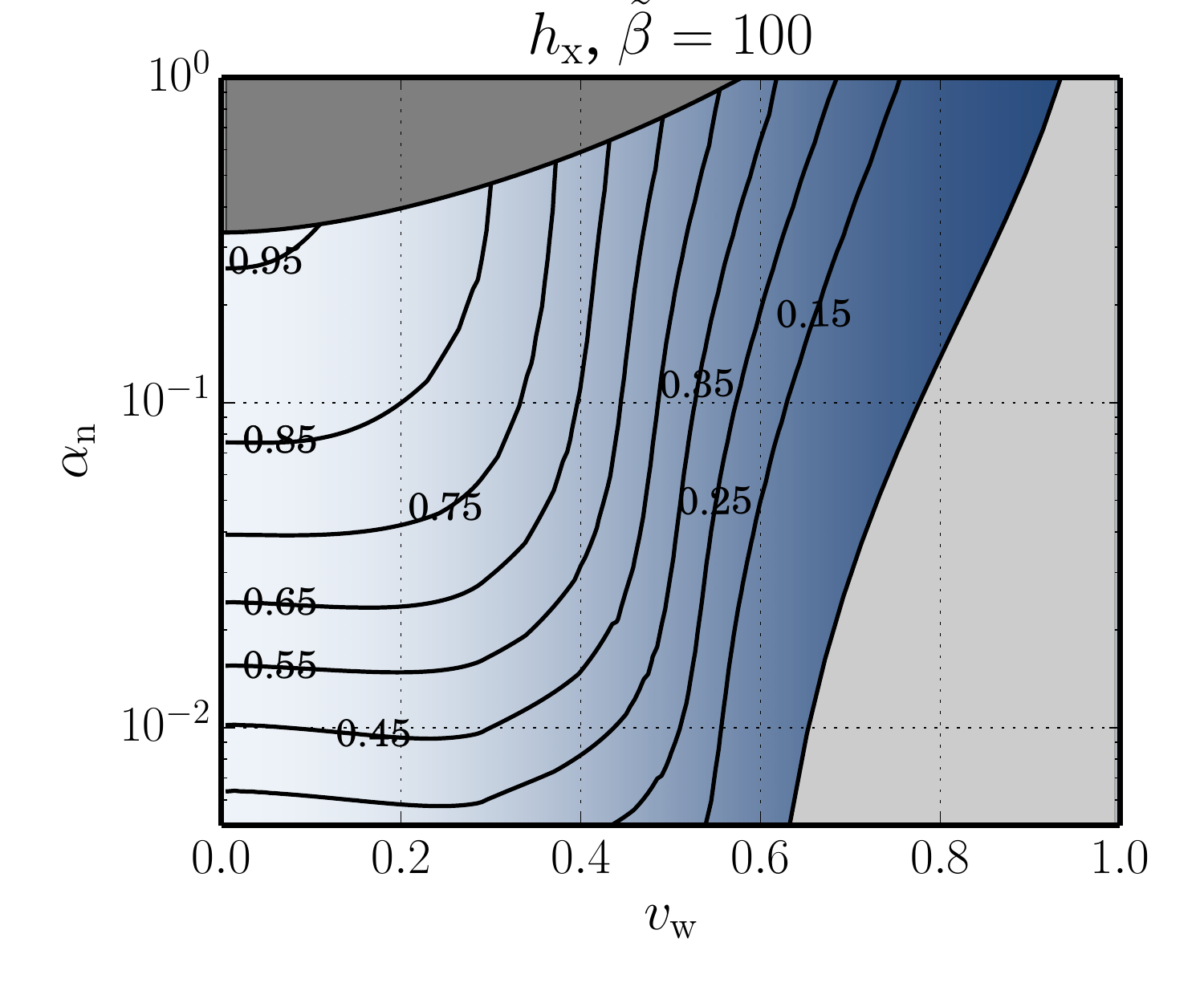}%\par\medskip
\includegraphics[width=0.33\textwidth]{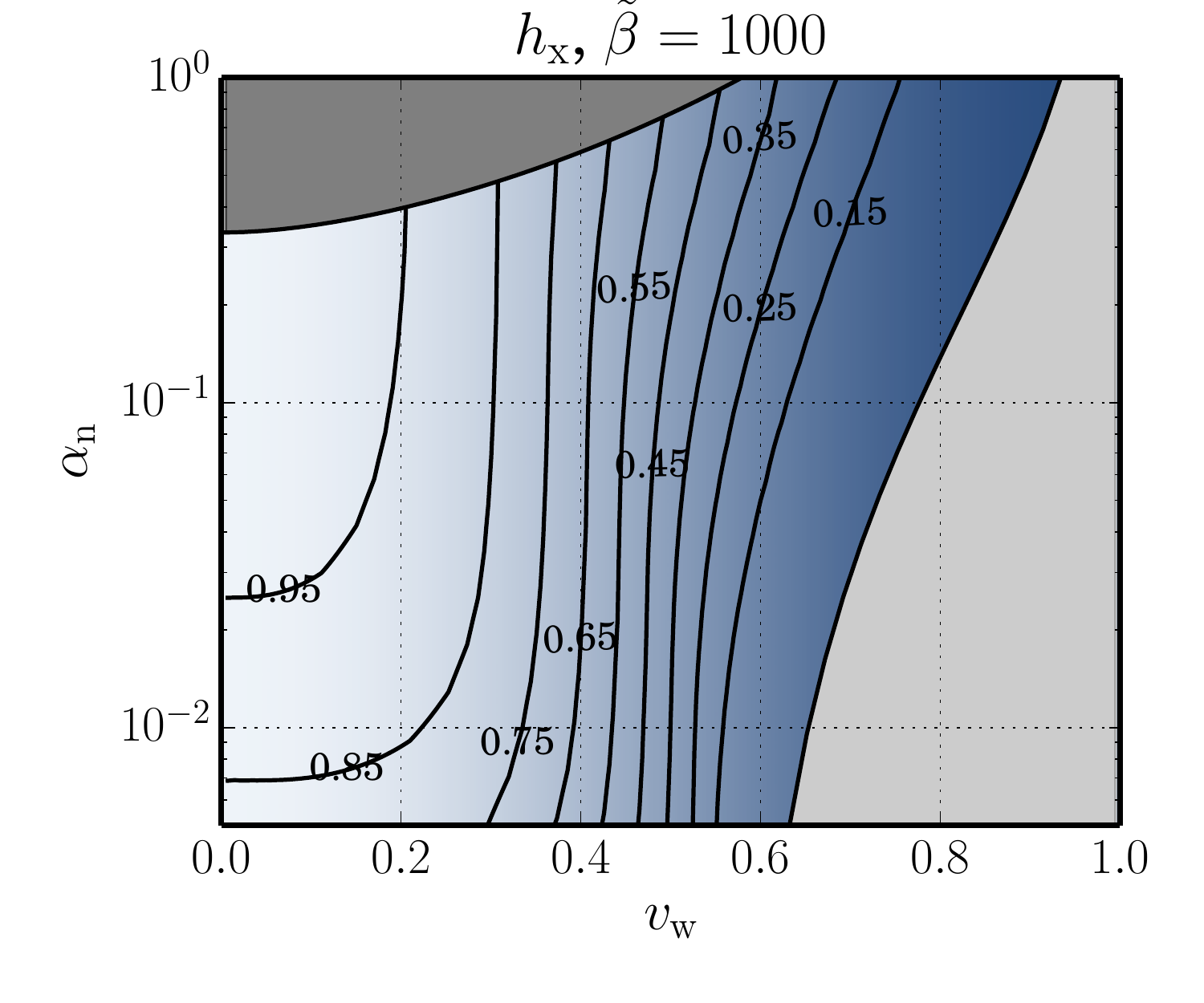}%\hfil\\
\caption{Top: contour plots of $v_{\rm eff}/v_{\rm sh}$, where $v_{\rm eff}$ is the expansion 
speed of the spherical shell inside which further bubble nucleation is effectively suppressed, and $v_{\rm sh}$ is the speed of the shock.  
For larger $\beta$, bubble nucleation is suppressed almost everywhere inside the shock, for a wide range of wall speeds $\vw$ and transition strengths $\strPar$. 
Bottom: contour plots of $\hstop$, the fractional volume occupied by the metastable phase at which bubble nucleation effectively stops, 
for values of 
the transition rate relative to the Hubble rate $\tilde\beta = 10, 100, 1000$ (left to right). 
As $\tilde\be$ increases the 
effect of nucleation suppression gets larger, due to the increasing   
sensitivity of the bubble nucleation rate to the temperature. 
In both rows, the blue shading shows the size of the maximum relative temperature change in the 
shell, with the same intensity map as in Fig.~\ref{contourtempdiff}. 
}
\label{contourhxfactor}
\end{figure*}

With this approximation in mind, we may write
\ben
%:
\EnhFac = \frac{3}{\vw^3}\int_{\vw}^{\vShock} \xi^2 ( 1 - e^{- \tilde\beta \De T(\xi)/\TN})  d\xi
\label{e:fEq}
\een
where $\tilde\beta = \beta/H$, $\TN$ is the nucleation temperature, and $\De T(\xi) = T(\xi) - \TN$.
We can rewrite the factor in terms of an effective wall speed $\vweff$,
\begin{equation}
    1+ f = \frac{\vweff^3}{\vw^3}.
\end{equation}
We plot the ratio $\vweff/\vsh$ on the top row of Fig.~\ref{contourhxfactor}.  We see that for fast ($\vw \gtrsim \cs$) bubble walls in strong ($\StrPar \gtrsim 0.3$) and rapid ($\tilde\be \gtrsim 100$) transitions, the estimate $\vweff \simeq \vsh$ is reasonably accurate.  However, for slower walls in weak and slow transitions, the approximation fails.

Now let us consider the situation with many bubbles. 
We label the bubbles by $i = 1,\ldots, \Nb$, and define a radial coordinate for each bubble $r_i$.  
Before bubbles start overlapping,  the effective volume in regions where bubble nucleation has stopped is 
\begin{align}
\Vbeff &=  \sum_{i=1}^{\Nb}  \frac{4\pi}{3} \vw^3(t' - t_i)^3 (1 + f) \nn\\
% &\times \left[ \frac{3}{\vw^3}\int_{\vw}^{\vShock} \xi_i^2 d\xi_i ( 1 - e^{-\De S}) \right].
&=  \sum_{i=1}^{\Nb}  \frac{4\pi}{3} \vweff^3(t' - t_i)^3 .
\end{align}
Conversely, the effective volume remaining where bubble nucleation can take place is 
\begin{equation}
    \Veff = \Vtot - \Vbeff .
\end{equation}
The rate of change of the fraction in the metastable phase is then modified to 
\begin{equation}
    \frac{dh}{dt} = -\vw h \frac{\Veff}{V} \int_{\tc}^{t} dt' p(t') 4\pi \vw^2(t - t')^2. 
\end{equation}
We take the ratio $\Veff/V$ to be that from a single bubble, \begin{equation}
    \frac{\Veff}{V} =     \frac{\Vtot - \Vbeff}{V} \simeq 
    \frac{(1+f)h - f}{h} .
\end{equation}
This approximation neglects regions outside 
the shells of radius $\vweff(t' - t_i)$ 
but inside $\vsh(t' - t_i)$, 
where the temperature in overlapping fluid shells has also reached a high enough value to suppress nucleation. 
This extra volume where nucleation is suppressed will give a small positive correction to $\vweff$.

We are also neglecting interactions between the shell of one bubble and the wall of a neighbouring one, which tend to slow down the expansion of the wall \cite{Cutting:2019zws}. We discuss this effect in the conclusions section.

We see that bubbles stop nucleating once $h$ drops below 
\ben
\hstop = \frac{f}{1+f} = 1 - \frac{\vw^3}{\vweff^3},
\een
so that $\hstop$ is the fractional volume at which the symmetric phase is reheated enough to prevent further bubble nucleation. 
% The inverse relation is $\EnhFac = \hstop/(1 - \hstop)$.
We plot $\hstop$ as a function of $\vw$ and $\strPar$, for three values of $\tilde{\beta}$, in the bottom row of Figure \ref{contourhxfactor}.

\begin{widetext}
Hence the equation for the fraction of the universe remaining in the metastable phase $h$ becomes 
\ben
% \begin{split}
\frac{dh}{dt} = -\vw h(t) (1+f) \int_{\tc}^{t} dt' p(t') \left[ 1  -  \frac{\hstop}{h(t')} \right] 
4\pi \vw^2(t - t')^2 \th(h(t') - \hstop), 
% \end{split}
\een
which upon integration with respect to $t$ 
becomes an integral equation, 
\ben
h(t) = \exp\left( -\frac{4\pi}{3} \vw^3 (1- \hstop)^{-1} \int_{\tc}^{t} dt' p(t') \left[ 1  -  \frac{\hstop}{h(t')} \right] (t - t')^3\th(h - \hstop)\right).
\label{hfun}
\een

With the approximation for the nucleation rate per unit volume (\ref{e:ProApp}),
and defining a dimensionless time variable 
$\tau=\beta t$, 
we can rewrite Eq.~(\ref{hfun}) as  
\begin{equation}
h(\tau) = \exp\left[-\frac{1}{6}(1- \hstop)^{-1}  \int_{\tau_c}^{\tau} d\tau^\prime  e^{\ta^\prime - \ta_f}
\left(1-\frac{h_x}{h(\tau^\prime)}\right)(\tau-\tau^\prime)^3\theta(h-h_x)\right]
\end{equation}
where we have used $p_f = \be^4/8\pi\vw^3$. 
We see that the equation for $h$ derived in the absence of the suppression effect (\ref{e:hDE}) is recovered in the limit $\hstop \to 0$. 
We assume that $ \ta_f - \ta_c \gg 1$, and hence that the solution depends very weakly on $\ta_c$.

We solve this equation for a given $\hstop$ by iteration:
\ben
h^{(a+1)}(\tau) = \exp\left[-\frac{1}{6}(1- \hstop)^{-1}  \int_{\tau_c}^{\tau} d\tau^\prime  e^{\ta^\prime - \ta_f}
\left(1-\frac{h_x}{h^{(a)}(\tau^\prime)}\right)(\tau-\tau^\prime)^3\theta(h-\hstop)\right]
\een
\end{widetext} 
starting with the solution at $f=0$, 
\ben
h^{(0)}(t) = \exp\left( -e^{\ta^\prime - \ta_f} \right). 
\een
The iteration converges very quickly and we stop after 5 iterations. The relative difference between the last two iterations depends on $\hstop$ and $\tau$ but is no greater than 0.01.

\begin{figure}[htbp]
\flushright
\includegraphics[width=0.48\textwidth]{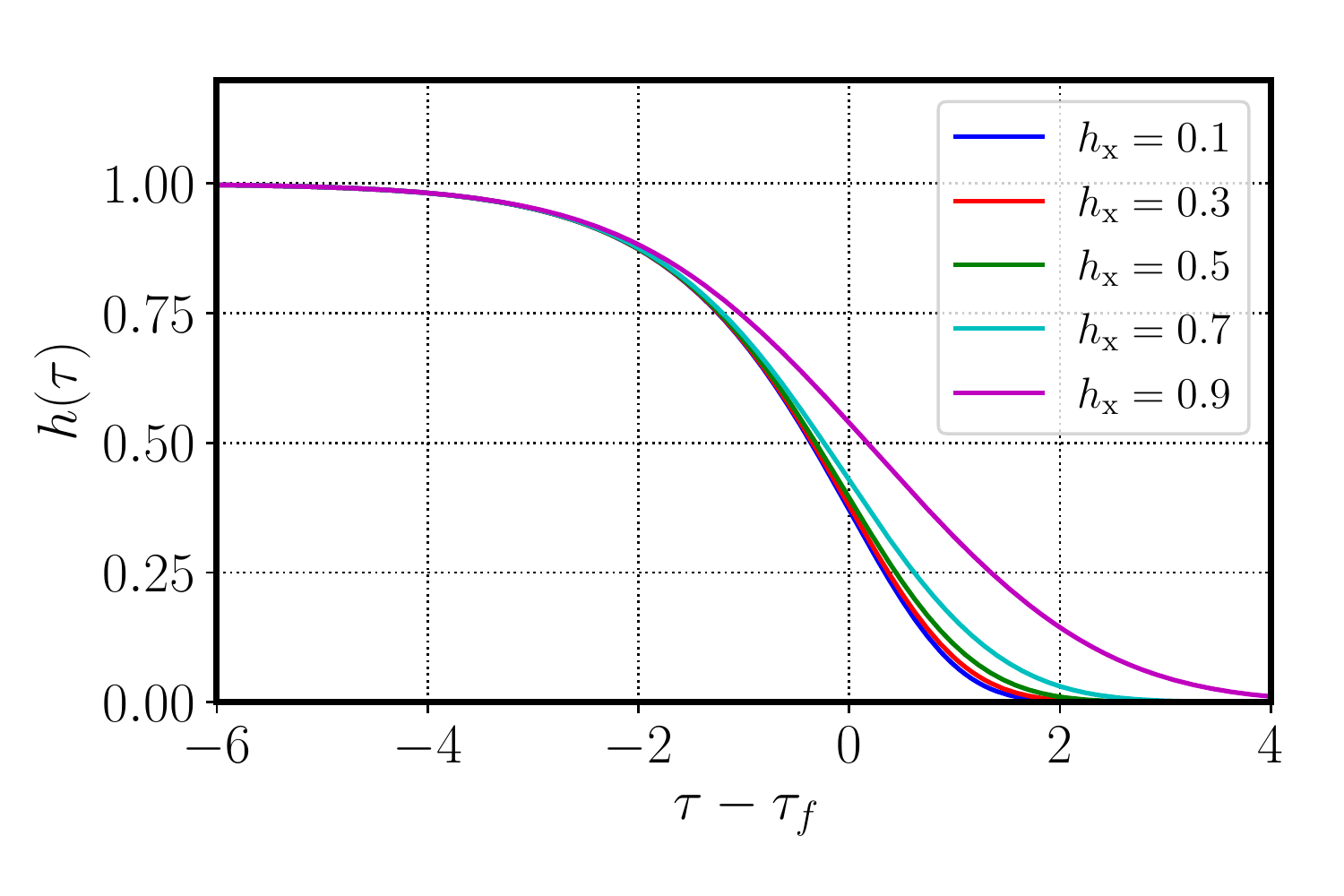}\\
\includegraphics[width=0.48\textwidth]{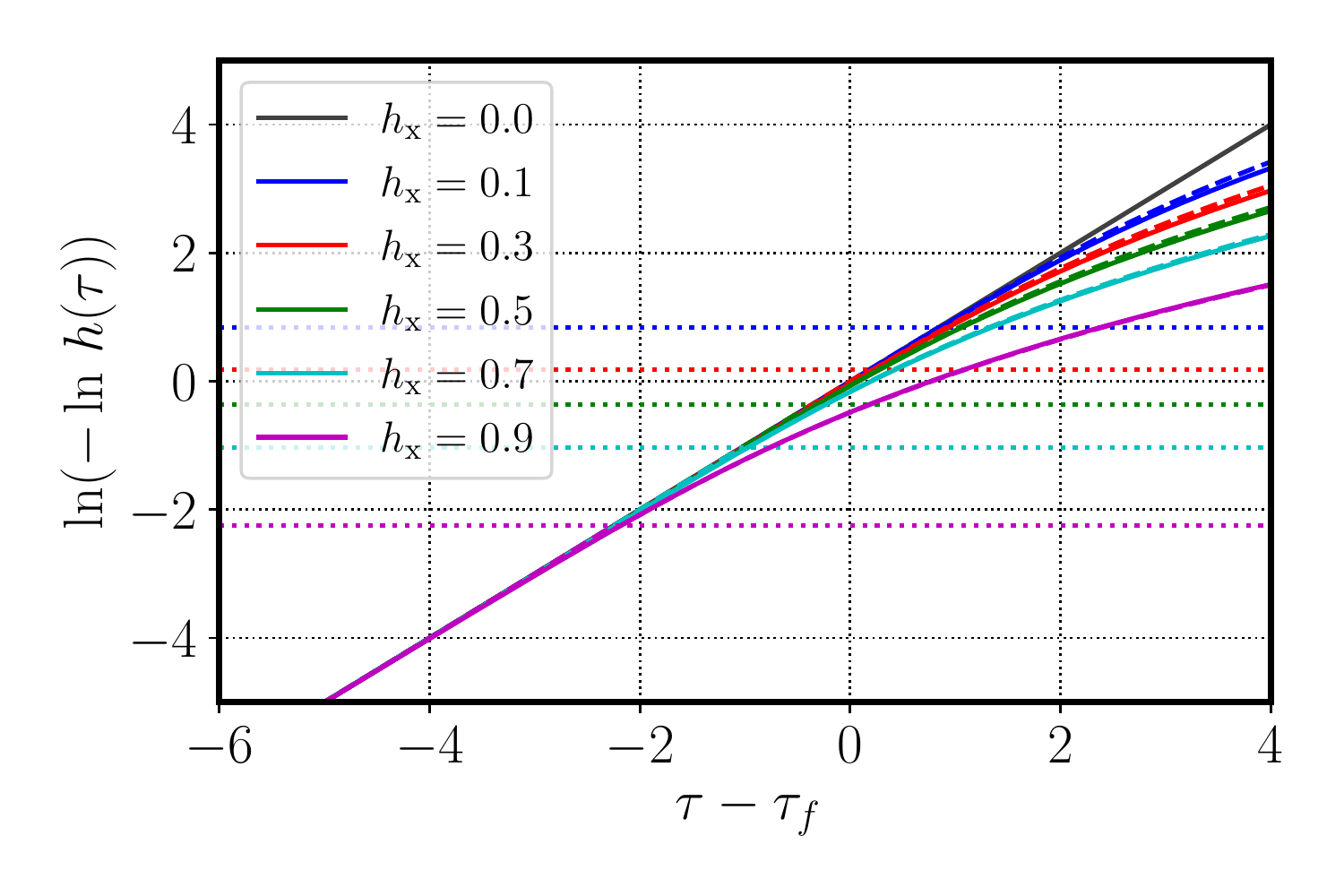}
\caption{%\label{fig:expTFH}
\small 
Plots of $h(\tau)$, where $h$ is the fraction of the universe in the metastable phase, for several values of the 
threshold fraction where nucleation stops $\hstop$. Solid lines are the numerical solution to Eq. \ref{hiter}, dashed lines (lower figure) 
are the approximation Eq. \ref{e:hAppSol}. Dotted lines give the threshold values $-\ln \hstop$.   
}
\label{htplot}
\end{figure}

For $f \gg 1$, $\hstop$ can be close to unity, i.e. bubble nucleation can effectively stop almost immediately. 
This can happen for large $\tilde{\beta}$. 

The fact that the iteration converges very fast motivates a simple approximation,
\begin{equation}
    h(\tau) \simeq \left\{
    \begin{array}{cc}
       h^{(0)}(\tau),  &  \tau \le \taustop \\
       h^{(1)}(\tau),  &  \tau > \taustop 
    \end{array}
    \right.
\end{equation}
Hence, for $\tau > \taustop$, 
\begin{widetext}
\ben
\label{hiter}
h(\tau) \simeq \exp\left( - \frac{1}{6} (1- \hstop)^{-1} \int_{\tau_c}^{\taustop} d\tau' e^{\tau' - \tau_f} \left(1-{\hstop}{e^{e^{\tau' - \tau_f} }}\right) (\tau - \tau')^3\right), 
\een
\end{widetext}
The integral can be performed by expanding the exponential (see appendix~\ref{happrox}), leading to the following equation:
\ben
\label{e:hAppSol}
h(\tau) \simeq \exp\left( -\frac{\estop}{6} \left(\la_0 \Delta\tau^3 + 3 \la_1 \Delta\tau^2 + 6 \la_2 \Delta\tau + 6 \la_3\ \right) \right)
\een
where $\estop \equiv \exp(\taustop - \tau_f)$ and 
\ben
\la_a = 1 -  \frac{\hstop}{1 - \hstop}\sum_{m=1}^\infty\frac{\estop^m}{(m+1)^a}
\een
To second order in $m$, 
\ben
\la_a \simeq 1 - \frac{\hstop}{1 - \hstop}\frac{\estop}{2^{a+1}} 
\een
At this order of approximation, $\estop$ can be solved exactly in terms of $\hstop$ through
\ben
h(\taustop) \equiv \hstop = \exp\left( - \estop  \la_3 \right),
\een
leading to the quadratic equation
\ben
- \frac{\hstop}{1 - \hstop}\frac{1}{16} \estop^2 + \estop + \ln \hstop = 0 .
\een
For $\hstop \to 0 $ we can neglect O($\hstop$) terms and the solution is $\estop = - \ln\hstop$, with 
\ben
h(\tau)  \simeq \exp\left( \frac{1}{6} \ln\hstop\left( \Delta\tau^3 + 3  \Delta\tau^2 + 6  \Delta\tau +6\ \right) \right) .
\een
For $\hstop \to 1 $, the second terms in the equation for $\la_a$ become important.  Writing $\hstop = 1 - \ep$, we have
\ben
\estop \simeq \ep, \quad \ep \simeq - \ln \hstop, \quad \la_a \simeq 1 - \frac{1}{2^{a+1}}. 
\een

\begin{figure}
\includegraphics[width=0.5\textwidth]{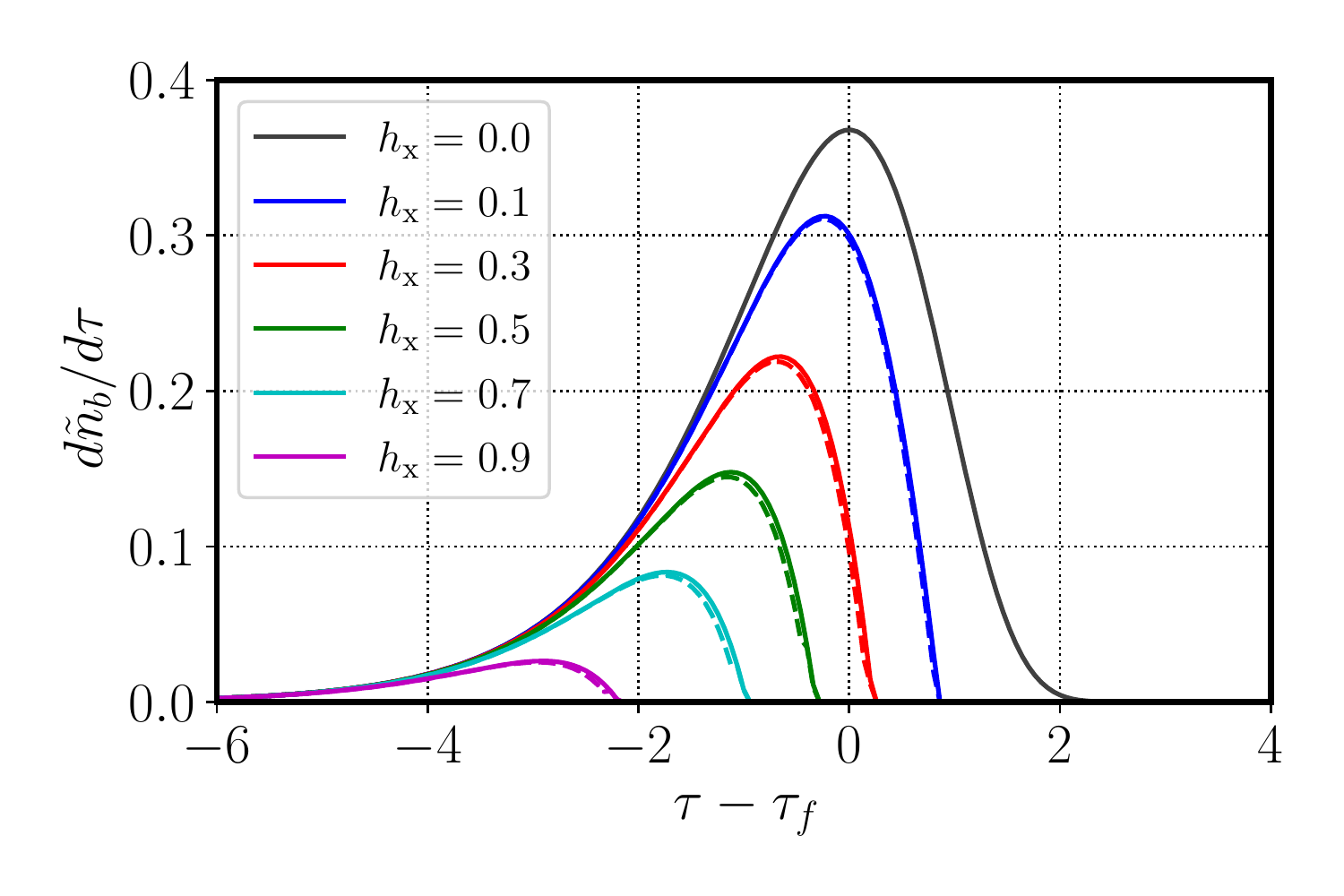}
\caption{\label{f:h_and_nb}
The universe-averaged dimensionless bubble nucleation rate where $\tilde n_b = n_b/n_b^{(0)}$,  the rate is given by   (\ref{e:BubDenDE}), and the reference bubble density is $n_b^{(0)} = \beta^3/8\pi \vw^3$
for the same values of $\hstop$ as Fig.~\ref{htplot}. 
}
\end{figure}

Figure~\ref{htplot} shows plots of $h(\tau)$ with different values of $\hstop$, obtained with the iterative method outlined above. 
It can be seem that for increasing $\hstop$, the transition takes longer, as a result of the reduced number density of bubbles nucleated. 
In the lower panel $\ln(-\ln(h))$ is plotted, along with the approximation derived above.

In the limit $\vw\to0$, the velocity in the fluid shell is small everywhere, and the approximate solution (\ref{e:delTEq}) can be used to estimate $f$.
Substituting Eq.~(\ref{e:delTEq}) in Eq.~\ref{e:fEq} 
yields 
\bea
\EnhFac &=& \frac{9 \alpha \tilde\beta}{2 \cs^2} \frac{  (1+\cs^2)}{(1 - 3\vw^2)^2} \int_{\vw}^{\cs}  \left( \xi - \frac{\xi^2}{\cs} \right) d\xi, \\
&=& \frac{3 \alpha \tilde\beta}{4 \cs^2} \frac{  (1+\cs^2)}{(1 - 3\vw^2)^2} \left( \frac{\vw^2(2\vw-3\cs)}{c_s} - \cs^2\right)
\label{e:fsmvEq}
\eea
where we have used the fact that the shock speed is approximately $\cs$ for deflagrations with low fluid speeds. This expression helps check numerical solutions at low $\vw$.

\section {Distance between bubbles}
In this section we will derive the equation for the distance between bubbles, for which we need to calculate the bubble number density.

To calculate the number of bubbles, 
we convert Eq.~(\ref{e:BubNumDen})
into an equation the bubble density $\nb = \Nb/\Vtot$,
\ben
\frac{d\nb}{dt} =  p(t)\left[ (1 + \EnhFac) h(t) -  \EnhFac \right] .
\label{e:BubDenDE}
\een
The density of bubbles is, on integrating (\ref{e:BubDenDE}), 
\begin{equation}
n_b = \frac{1}{1-\hstop}\int_{t_c}^{\tstop} p(t) \left( h(t) - \hstop\right) dt ,
\end{equation}
where $\tstop$ is the time at which nucleation stops, i.e.~where $h = \hstop$.

Introducing the function 
\begin{equation}
I_h(\hstop) = \frac{1}{1-\hstop}\int_{\tau_c}^{\taustop} e^{\tau - \ta_f} \left( h(\ta) -\hstop\right) d\ta 
\end{equation}
we have that 
\begin{equation}
n_b(\hstop) = \be^{-1}{p}_f I_h(\hstop) = \nb^{(0)}  I_h(\hstop),
\end{equation}
where $\nb^{(0)} = 8\pi \vw^3 \beta^3$ is the bubble density in the absence of nucleation suppression. 
In general, we expect $0 < I_h(\hstop) < 1$.  This function represents the reduction in the mean bubble density 
by the suppression of the nucleation in advance of the bubbles wall. 
Clearly, $I_h(0) = 1$.
We plot the bubble nucleation rate from the numerical solutions, and the analytic approximation, 
in Fig.~\ref{f:h_and_nb}. 

We recall that the mean bubble centre spacing is defined as 
$\Rbc = \nb^{-1/3}$. Hence, the mean bubble centre spacing is increased by a 
factor 
\ben
\Lambda(\hstop)\equiv \frac{\Rbc}{\Rbc{(0)}} = I_h^{-1/3}(\hstop)  ,
\een
where $R_*(0) = (8\pi)^{1/3}\vw/\beta$ is the mean spacing in the 
absence of nucleation suppression.  
Normalised this way, we have $\Lambda(0) = 1$.

\begin{figure}[htbp]
\centering
\includegraphics[width=0.4\textwidth]{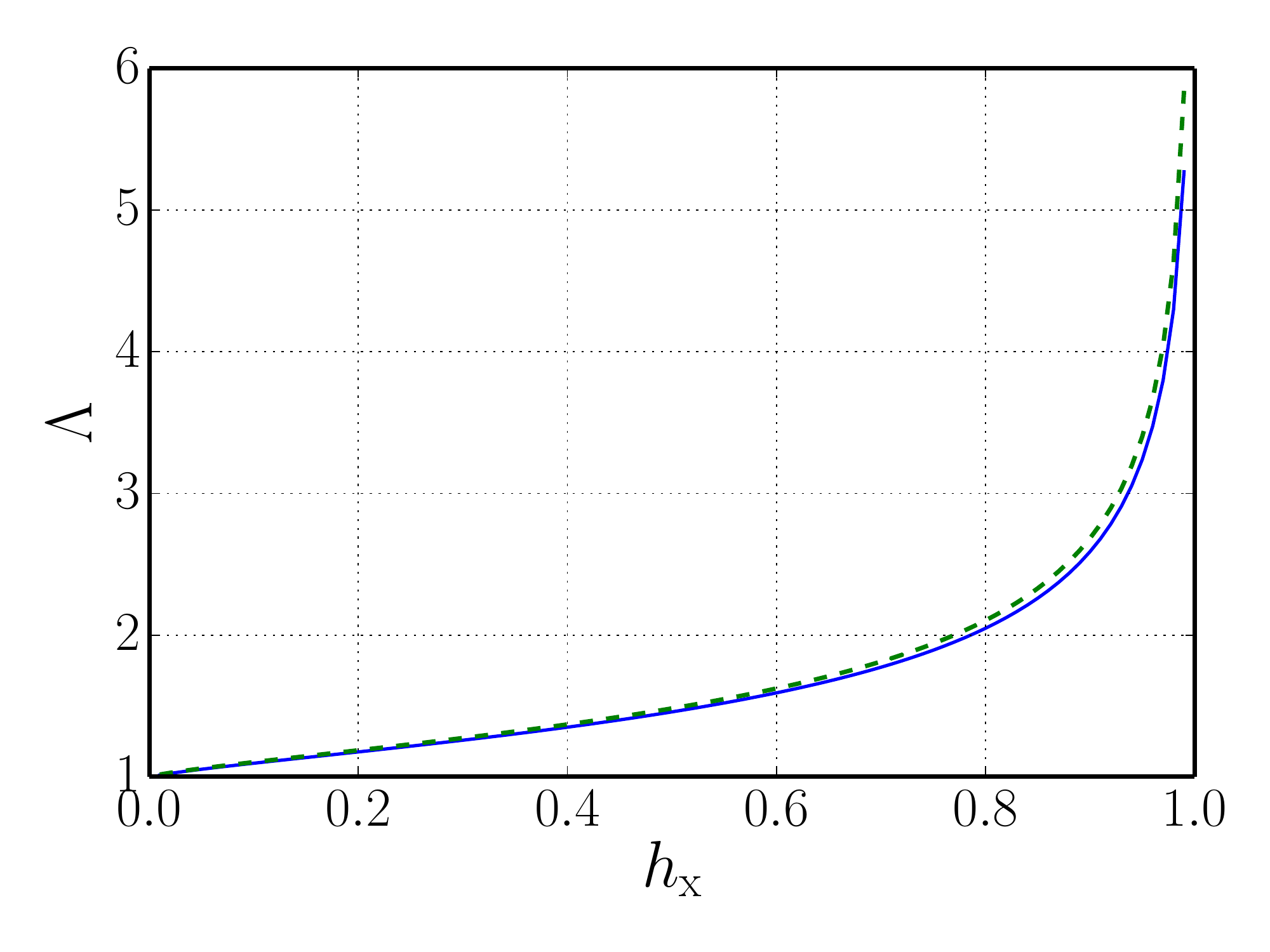}
\vskip 0.1 cm
\caption{Bubble spacing enhancement factor $\Lambda$ as a function of 
the fractional volume of the universe occupied by the metastable phase at which bubble nucleation stops, $\hstop$. The blue line uses the numerical solution and the dashed line uses the analytic approximation. 
As $\hstop\to1$, bubble nucleation stops earlier, and the bubbles that are nucleated grow to larger sizes. 
}
\label{chx}
\end{figure}

The integral can be performed with $h(\tau)$ in its approximate form, leading to 
\begin{equation}
\label{e:AnaAppIh}
    I_h(\hstop) = 1 + \frac{\hstop \ln \hstop}{1 - \hstop}.
\end{equation}
Figure~\ref{chx} shows the bubble spacing enlargement factor  $\Lambda(\hstop)$, computed from the numerical solutions for $h(\tau)$, 
along with the analytic approximation calculated from \eqref{e:AnaAppIh}.
As $\hstop$ increases $\Lambda(\hstop)$ increases demonstrating that 

\begin{figure*}[htb!]
\centering
\includegraphics[width=0.33\textwidth]{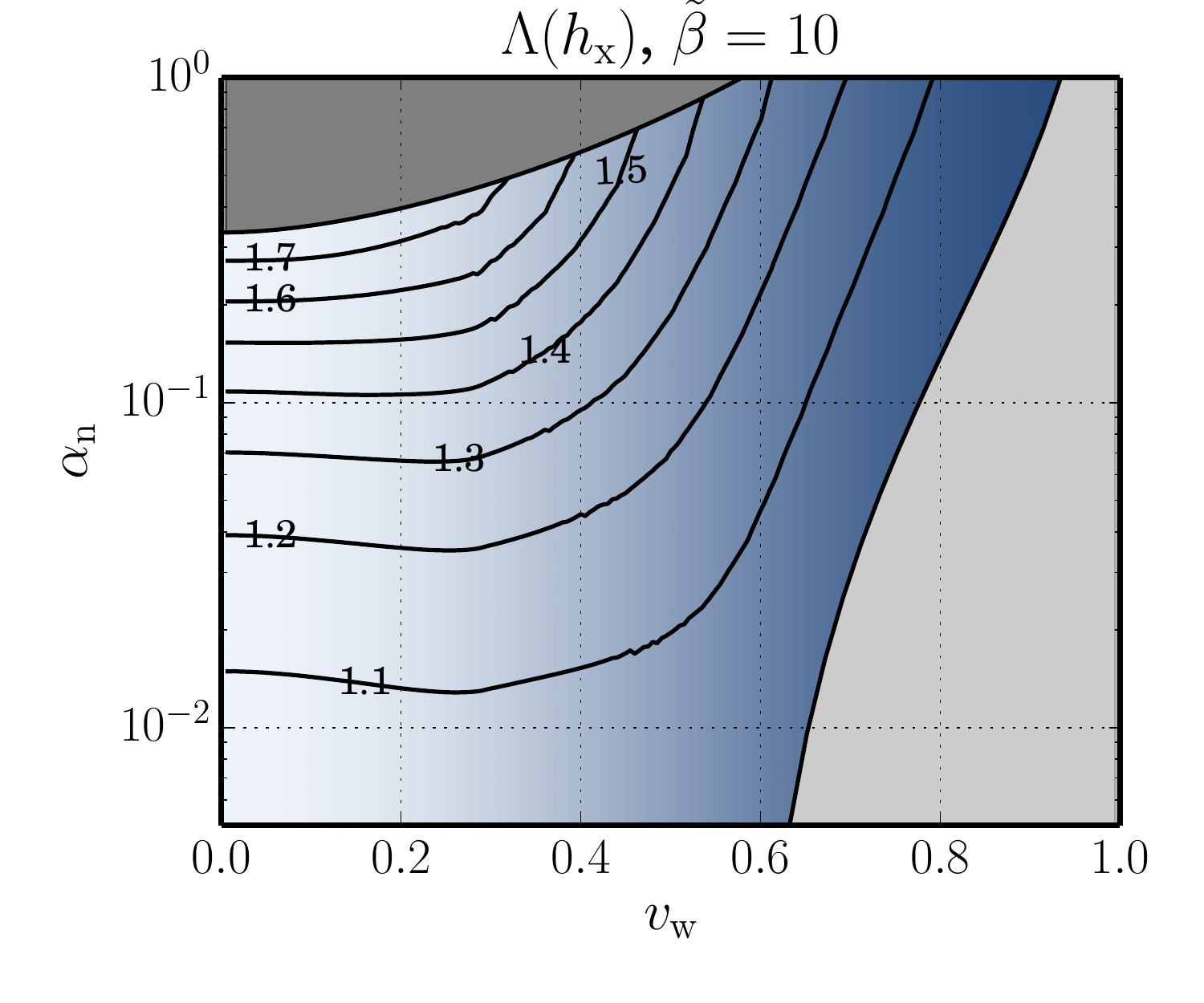}%\hfil
\includegraphics[width=0.33\textwidth]{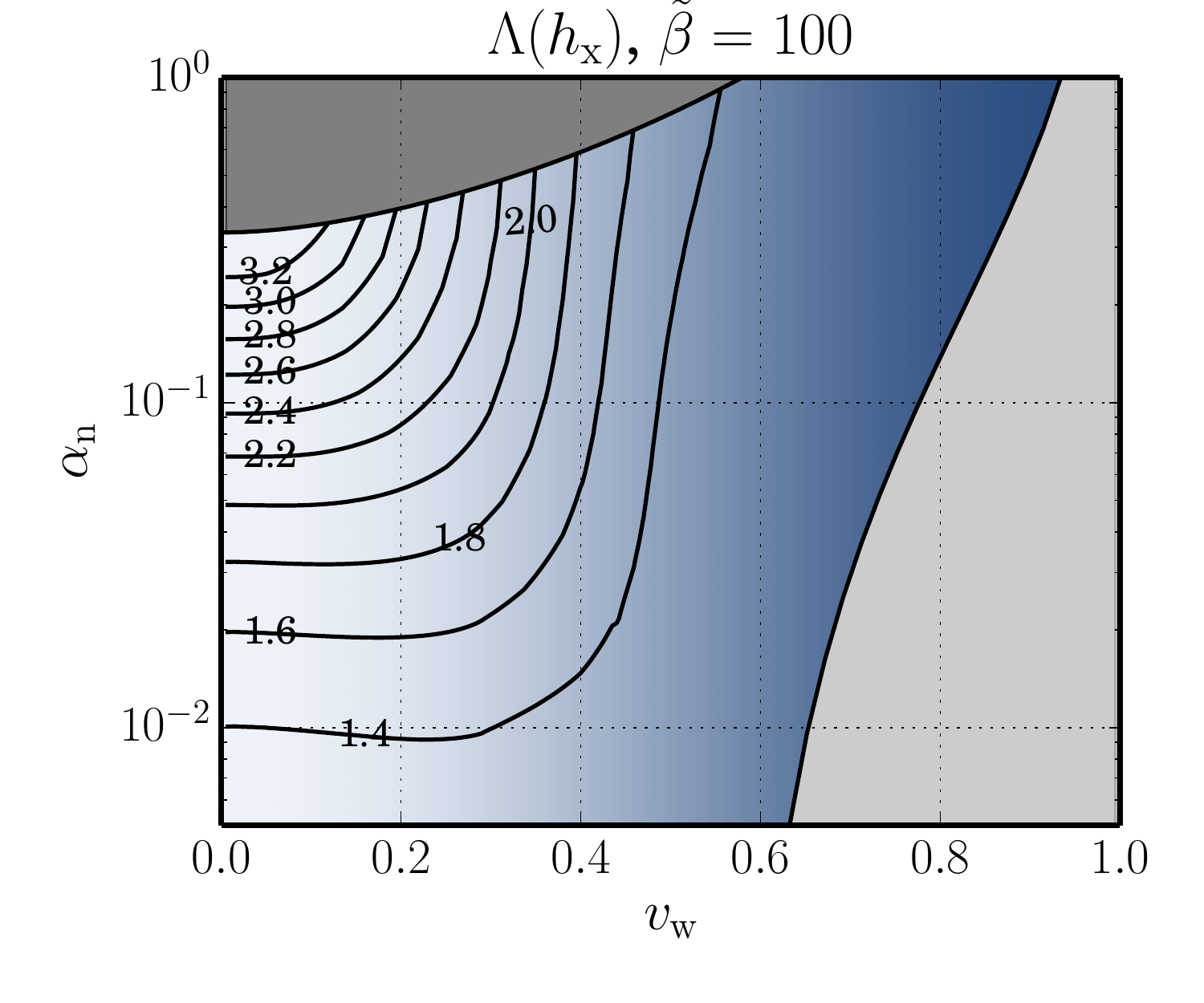}%\par\medskip
\includegraphics[width=0.33\textwidth]{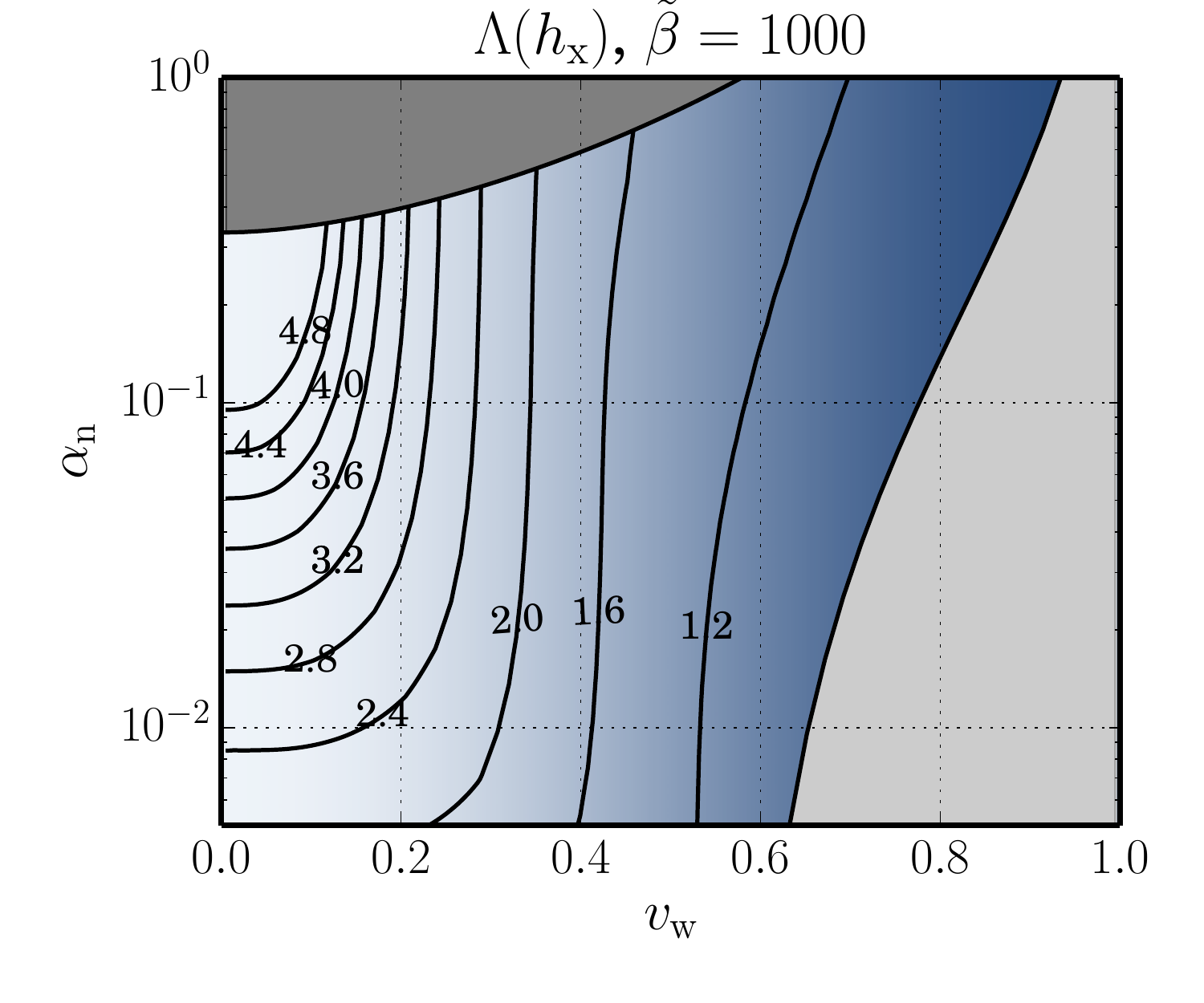}\\%\hfil
% \end{figure*}
% \begin{figure*}[htb!]
% \centering
\includegraphics[width=0.33\textwidth]{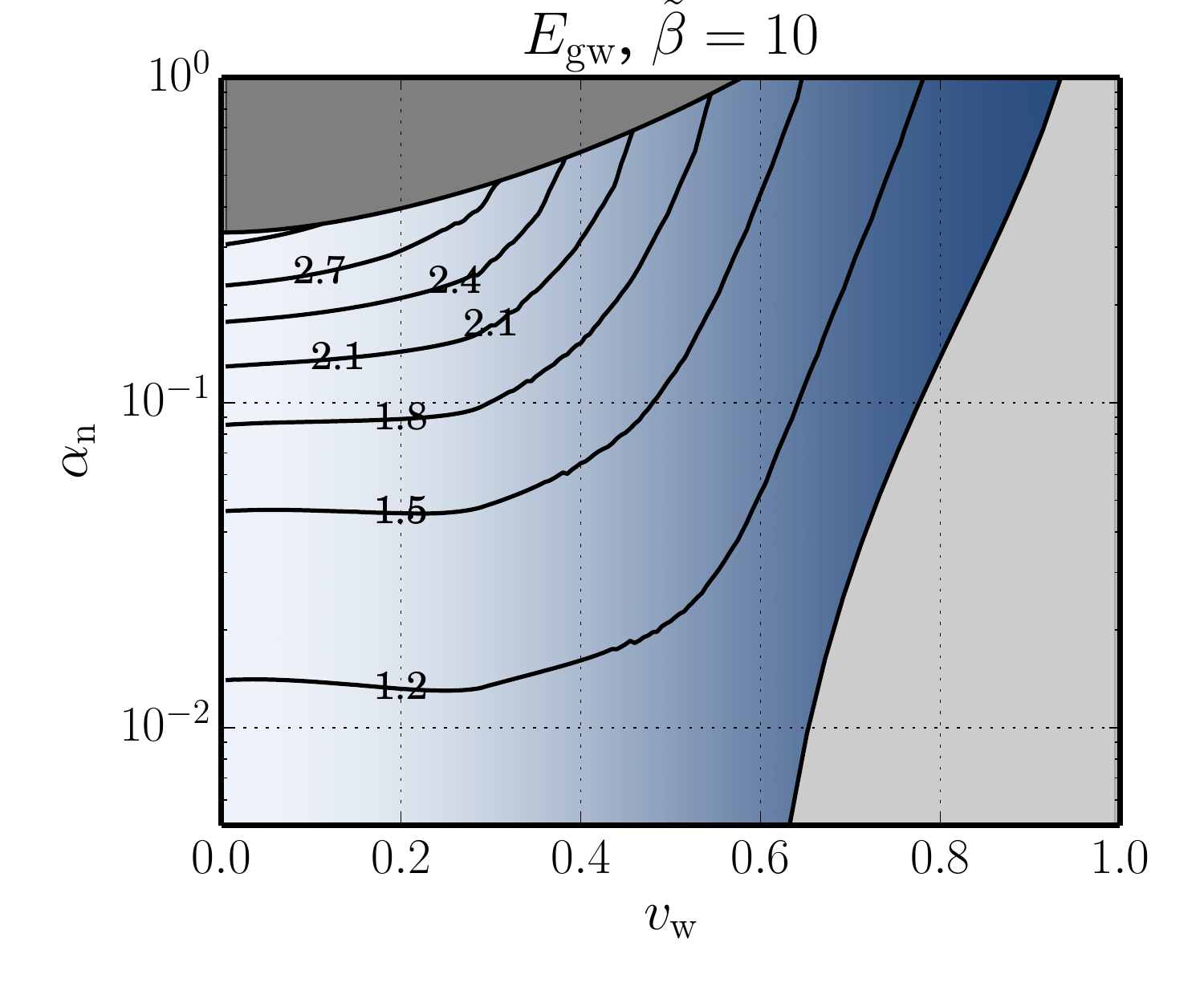}%\hfil
\includegraphics[width=0.33\textwidth]{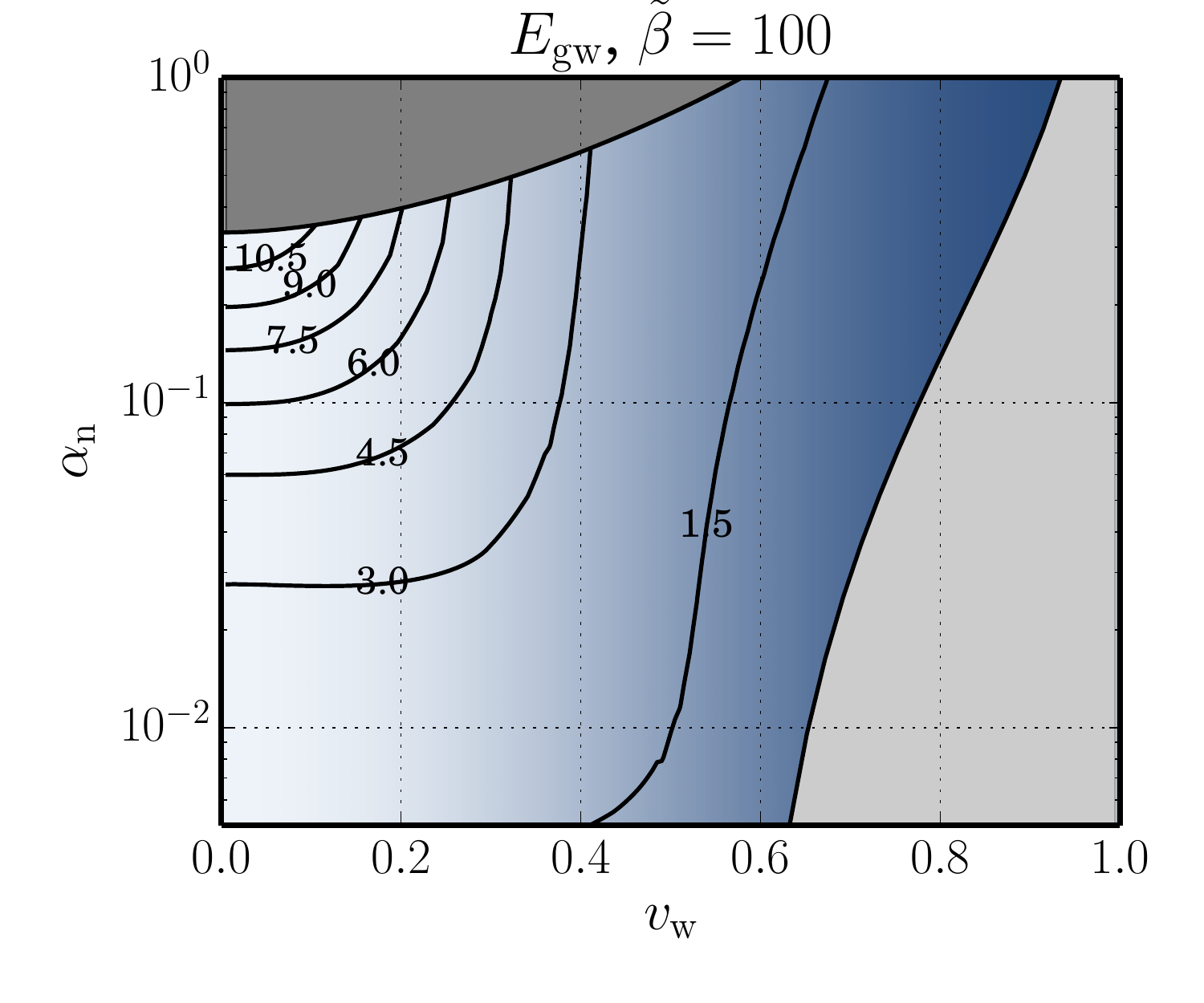}%\par\medskip
\includegraphics[width=0.33\textwidth]{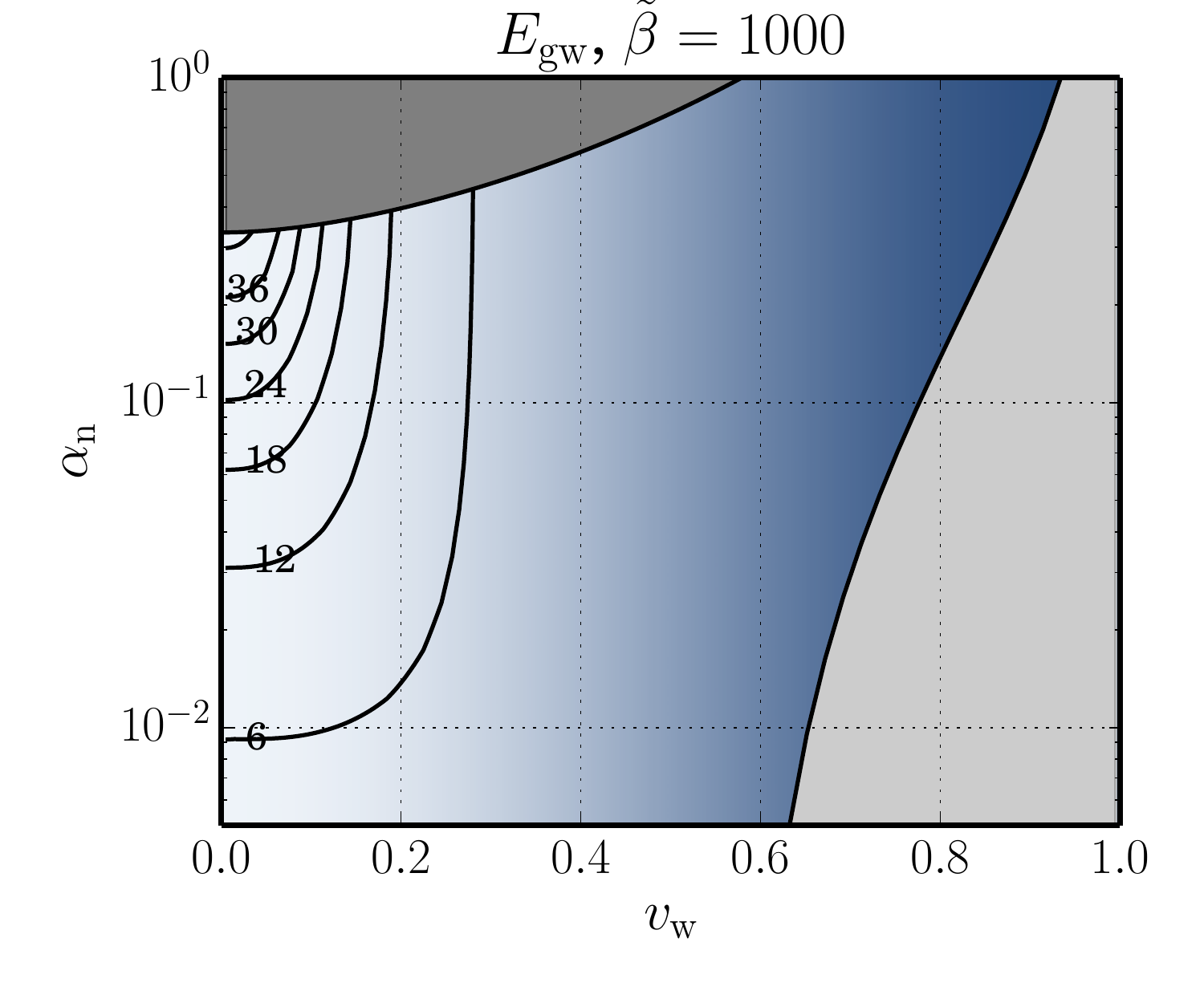}%\hfil
\caption{Top: Contour plots of the bubble size enhancement factor 
$\Lambda$ ratio in the plane of wall speed $\vw$ and 
transition strength parameter $\strPar$, 
for ratios of the transition rate to Hubble rate $\tilde\beta=10, 100, 1000$ (left to right).
Contour plots of GW enhancement factor  $E_\text{gw}$ for different  temperatures in the plane of $\alpha_n$ and $\xi_w$. As $\tilde\beta$ (the transition rate) increases the $E_\text{gw})$  increases. 
In both rows, the blue shading shows the size of the maximum relative temperature change in the 
shell, with the same intensity map as in Fig.~\ref{contourtempdiff}.} 
\label{contourGWfactor}
\end{figure*}

Figure \ref{contourGWfactor} (top row) shows contour plots of $\Lambda(\hstop)$ in the plane of wall speed $\vw$ and transition strength parameter $\strPar$, for $\tilde\beta=10, 100, 1000$. The detonation region and hydrodynamically inaccessible values are greyed out. 

\section{Gravitational wave power}

The gravitational wave power spectrum produced by a first order phase transition is, in a large region of parameter space, dominated by 
acoustic production \cite{Hindmarsh:2013xza,Hindmarsh:2015qta,Hindmarsh:2017gnf,Hindmarsh:2019phv} (see also \cite{Weir:2017wfa,Mazumdar:2018dfl,Caprini:2019egz,Hindmarsh:2020hop} for reviews).
The total gravitational wave power is, provided that the mean bubble size is much less than the Hubble length, 
 \ben\label{Eq:Omgw_ssm}
 \OmGW = 3K^{2}(\vw,\al)\left(\HN \tau_{\mathrm{v}}\right)\left(\HN R_*\right) \tilde\Omega_{\rm gw},
 \een
where $K$ is the fraction of the energy of the fluid in the form of kinetic energy, 
$\HN$ is the Hubble rate at nucleation (assumed to be the same as the Hubble rate at the end of the transition),
$\tLife$ is the effective lifetime of the source, and $\tilde\Omega_{\rm gw} \simeq 10^{-2}$ is a dimensionless parameter 
characterising the efficiency of gravitational wave production. 

The effective source lifetime is the shorter of the Hubble time and the shock appearance timescale 
$\tShock = R_*/\sqrt{K}$: once shocks appear, the kinetic energy is dissipated in a time of order a few $\tShock$. 
An investigation of how a shear stress source is diluted by expansion \cite{Hindmarsh:2015qta,Guo:2020grp} shows that 
to a first approximation, in which the source is constant and shuts off after time $\tShock$,
 \ben{\label{Eq:Hntv}}
\HN \tau_v \simeq \left(1 -  \frac{1}{\sqrt{1 + 2x}} \right).
 \een
For convenience we define
\ben
\label{Eq:scaling_factors_GW}
J = \HN R_* \HN\tau_v  = r_* \left(1 -  \frac{1}{\sqrt{1 + 2x}} \right),
\een
where $r_* =  \HN R_*$ and $x = r_* / \sqrt{K} $. 
%with $K$ the kinetic energy fraction of the bubble. 
Recalling the definition of the 
bubble spacing enhancement factor $\Lambda$, 
the GW power is also enhanced by a factor 
\begin{eqnarray}
E_\text{gw}(\vw,\al,\tilde{\be}) &=& \Lambda \left(1 -  \frac{1}{\sqrt{1 + 2\Lambda r_*(0)/K^{1/2} }} \right).
\end{eqnarray}
where the Hubble-scaled mean bubble spacing without nucleation suppression is \cite{Enqvist:1991xw} 
\ben
r_*(0) =  \Rbc{(0)} \Hn. 
\een
In Fig.~\ref{contourGWfactor} (bottom row) we show contour plots of the GW enhancement factor $E_\text{gw}$
for our standard values $\beta/\HN = 10, 100, 1000$.
The kinetic energy fraction $K$ has been evaluated using the single-bubble kinetic energy fraction
\begin{equation}
    K = \frac{3}{\vw^3 e_\text{n}} \int d\xi  \xi^2 w\ga^2 v^2,
\end{equation}
where $v$ and $w$ are the solution to Eqs.~(\ref{e:SelSimVel},\ref{e:SelSimVel}), and $e_\text{n}$ is the energy density outside the expanding fluid shell. The kinetic energy density is calculated from the numerical solutions, integrated using the trapezium rule.

\section{Conclusions}

In this paper we have studied  the suppression of bubble nucleation in cosmological phase transitions proceeding by deflagrations. In a deflagration, 
some of the energy released by the transition goes into heating up the fluid in front of the bubble wall which, as a result, suppresses further bubble nucleation. In a detonation, on the other hand, the bubble wall is ahead of the shell of excess thermal energy, and the effect is absent.

We find that nucleation stops when a certain fraction of the volume in the metastable phase has been converted.  The fraction can easily be computed from the solution of the relativistic hydrodynamic equations, in an expansion in the relative temperature fluctuation $ \De T/\Tn$. We solve the equations for a fluid with a bag model equation of state, and compute the first order effect. This is sufficient for transition strengths below around $0.3$, and wall speeds below the sound speed.

The suppression of nucleation results in a lower number of bubbles per unit volume, and therefore a larger mean distance between their centres. The effect results in a larger intensity of gravitational waves from the transition.

The region of higher temperature extends outward to a leading shock, which travels faster than the sound speed.  For this reason it has sometimes been estimated that the region extends out to the shock speed $\vsh$ \cite{Moore:2000jw} or the sound speed (an estimate of the shock speed) \cite{Caprini:2019egz}. Here we have shown that the effect is more complicated. 
The suppression can be expressed as the effective speed $\vweff$ of expanding spherical volumes inside which nucleation stops, with $\vw < \vweff < \vsh$.  We show that this approximation $ \vweff \simeq \vsh$ works well for fast walls in strong and rapid transitions, but not otherwise. 

The more rapid the phase transition, as measured by the parameter $\tilde\beta$, the more sensitive the system is to the suppression effect.  This is because  $\tilde\beta$ is equal to the logarithmic derivative of the nucleation probability with respect to the temperature.  Increasing the phase transition strength parameter $\strPar$ also increases the effect, as one would expect from the larger release of thermal energy.  The effect also increases with decreasing wall speed $\vw$, as the heated volume is larger relative to the bubble size.

For example, for ($\vw$, $\strPar$, $\tilde\be$) = (0.1, $3\times 10^{-2}$, $1000$), 
the ratio $ \vweff/\vsh \simeq 0.4$, and bubbles stop nucleating when only 5\% of the universe has been converted to the stable phase.  This has the effect of increasing the mean bubble spacing by a factor 4, and the gravitational wave intensity by a factor 5. We show the magnitude of both effects, as functions of $\vw$ and $\strPar$, in contour plots in Fig.~\ref{contourGWfactor}.

Our results are derived from a numerical solution to an integral equation for the fraction remaining in the metastable phase as a function of time, $h(t)$. We have also shown that good numerical approximations exist, and that the suppression factors can be calculated from the solution of the relativistic hydrodynamic equations.

A further effect to consider for precise calculations of the gravitational wave power spectrum is the altered collision time distribution \cite{Hindmarsh:2019phv}.  In the standard calculation with exponentially growing nucleation rate per unit volume of metastable phase, the distribution of times between a segment of wall being nucleated, and colliding with another segment of wall, is distributed exponentially.  If all bubbles are nucleated simultaneously, the distribution is a power times an exponential. As $\hstop$ is reduced from 1 to 0, we are effectively interpolating between these two situations, and we therefore expect the shape of the gravitational wave power spectrum to interpolate between the exponential and simultaneous \cite{Hindmarsh:2019phv} as well. 

Finally, in this paper we have assumed that the walls expand with a constant speed throughout the transition.  On the other hand, when a bubble wall encounters the heated region surrounding another bubble, the pressure difference across it will be reduced, and the wall will slow down \cite{Cutting:2019zws}.  If the nucleation has effectively stopped by the time the bubble walls start to slow, the number of bubbles nucleated per unit volume, and hence the mean bubble spacing $\Rbc$, will not be affected.  The effect of the walls slowing will therefore be smaller for larger $\hstop$, and hence larger $\tilde\beta$.  
%The nucleation rate is exponentially sensitive to  small temperature changes, but the pressure is only linearly dependent.  
We therefore expect the slowing of the walls to be important only for lower values of $\tilde\beta$. We will explore the effect in more detail elsewhere.

\begin{acknowledgments}
We thank Daniel Cutting, Stephan Huber, Jose-Miguel No, Kari Rummukainen and David Weir for useful comments and discussions. M Al-Ajmi (ORCID ID 0000-0001-9888-5318) acknowledges the University of Sussex for hosting him in his sabbatical leave period. He also thanks Sultan Qaboos University for support with grants IG/SCI/PHYS/17/05 and IG/SCI/PHYS/20/02. M.H. (ORCID ID 0000-0002-9307-437X) acknowledges support from the Academy of Finland grant no.~333609. 
Computations were performed using NumPy \cite{2020NumPy-Array} and SciPy \cite{2020SciPy-NMeth}. All plots were made with Matplotlib \cite{Hunter:2007}.
\end{acknowledgments}

\appendix

\section{Detailed calculation of $\hstop$}
\label{happrox}
We study the integral in the exponent of Eq.~\ref{hiter},
\bea
L(\tau) &=& \int_{\tau_c}^{\taustop} d\tau' e^{\tau' - \tau_f} \left(1-{\hstop}{e^{e^{\tau' - \tau_f} }}\right) (\tau - \tau')^3.
\eea
By expanding the first exponential, and writing $\estop \equiv e^{\tau' - \tau_f}$
we have
\bea
% \int_{\tau_c}^{\taustop} d\tau' e^{\tau' - \tau_f} \left(1-{\hstop}\sum_{m=0}^\infty \frac{e^{m(\tau' - \tau_f)}}{m!} \right) (\tau - \tau')^3 \nonumber\\
L(\tau)  &=&  e^{\taustop - \tau_f}\int_{\tau_c}^{\taustop} d\tau' e^{\tau' - \taustop} \left(1-{\hstop}\sum_{m=0}^\infty \frac{\estop^{m}}{m!} e^{m(\tau' - \taustop)}\right) \nonumber \\
 &\times&(\Delta\tau  + \taustop -  \tau')^3, \nonumber\\
\eea
where $\Delta\tau = \tau - \taustop$.
As the integrals are dominated by their upper limits, it is a good approximation to take $\tau_c \to -\infty$, leading us to consider
\bea
K_m(\tau) &=& \int_{-\infty}^{\taustop} d\tau' e^{(m+1)(\tau' - \taustop)} (\Delta\tau  + \taustop -  \tau')^3 \nonumber\\
&=& k_m^0 \Delta\tau^3 + 3 k_m^1 \Delta\tau^2 + 3 k_m^2 \Delta\tau + k_m^3  ,
\eea
where
\bea
k_m^a 
% &=& \int^{\taustop}_{-\infty} d\tau'  e^{(m+1)(\tau' - \taustop)} \left( \taustop -  \tau'  \right)^a \nonumber\\
% &=& \int_0^\infty dx e^{-(m+1)x} x^a \nonumber\\
&=& \frac{a!}{(m+1)^a}
\eea
Hence
\bea
L(\tau) &=& (1 - \hstop)\sum_{m=0}^\infty {A_m} \left(k_m^0 \Delta\tau^3 + 3 k_m^1 \Delta\tau^2 + 3 k_m^2 \Delta\tau + k_m^3\right)
\nonumber
% &=& (1 - \hstop)\sum_{m=0}^\infty {A_m}  \left(\frac{1}{m+1} \Delta\tau^3 + \frac{3}{(m+1)^2}  \Delta\tau^2\right. \nonumber \\
% &+& \left. \frac{6}{(m+1)^3}  \Delta\tau +  \frac{6}{(m+1)^4}\right)
% \nonumber\\
\eea
where
\ben
A_m = \left\{
\ba{cc}
1, & m=0, \\[6pt]
\displaystyle - \frac{\hstop}{1 - \hstop} \frac{e^{m(\taustop - \tau_f)}}{m!}, & m > 0. 
\ea
\right.
\een
Finally, we write  
\ben
\label{e:hAppSolApp}
h(\tau) \simeq \exp\left( -\frac{\estop}{6} \left(\la_0 \Delta\tau^3 + 3 \la_1 \Delta\tau^2 + 6 \la_2 \Delta\tau + 6 \la_3\ \right) \right),
\een
where 
\ben
\la_a = 1 -  \frac{\hstop}{1 - \hstop}\sum_{m=1}^\infty\frac{\estop^m}{(m+1)^a}.
\een

\bibliography{GWs}

\end{document}